\documentclass[aps,pre,preprint,onecolumn,superscriptaddress,floatfix]{revtex4}

\pdfoutput=1

\usepackage{graphicx}
\usepackage{amssymb,amsmath}
\usepackage{color}
\usepackage{bm}
\usepackage{bbm}
\usepackage{setspace}

\newcommand{\R}{\mathbb{R}}
\newcommand{\norm}[1]{\left|\left|#1\right|\right|}

\newcommand{\trace}[1]{\mathrm{trace}\left( #1 \right)}

\begin{document}

\singlespacing

\title{Interest communities and flow roles in directed networks: the
  Twitter network of the UK riots}

\author{Mariano Beguerisse-D\'iaz} 
\email{m.beguerisse@imperial.ac.uk}
\affiliation{Department of
  Mathematics, Imperial College London, London SW7 2AZ, UK.}
\affiliation{Department of  Chemistry, 
Imperial College London, London SW7 2AZ, UK.} 
\author{Guillermo  Gardu\~no-Hern\'andez}
\affiliation{Sinnia, Mexico City, Mexico} 
\author{Borislav Vangelov}
\affiliation{Department of  Mathematics, 
Imperial College London, London SW7 2AZ, UK.} 
\author{Sophia~N. Yaliraki}
\affiliation{Department of  Chemistry, 
Imperial College London, London SW7 2AZ, UK.} 
\author{Mauricio Barahona}\email{m.barahona@imperial.ac.uk}
\affiliation{Department of  Mathematics, 
Imperial College London, London SW7 2AZ, UK.} 


\begin{abstract}
Directionality is a crucial ingredient in many complex networks in
 which information, energy or influence are transmitted. In such
 directed networks, analysing flows (and not only the strength of
 connections) is crucial to reveal important features of the network
 that might go undetected if the orientation of connections is
 ignored.  We showcase here a flow-based approach for community
 detection in networks through the study of the network of the most
 influential Twitter users during the 2011 riots in England.  Firstly,
 we use directed Markov Stability to extract descriptions of the
 network at different levels of coarseness in terms of interest
 communities, i.e., groups of nodes within which flows of information
 are contained and reinforced. Such interest communities reveal user
 groupings according to location, profession, employer, and topic.
 The study of flows also allows us to generate an interest distance,
 which affords a personalised view of the attention in the network as
 viewed from the vantage point of any given user.  Secondly, we
 analyse the profiles of incoming and outgoing long-range flows with a
 combined approach of role-based similarity and the novel relaxed
 minimum spanning tree algorithm to reveal that the users in the
 network can be classified into five roles. These flow roles go beyond
 the standard leader/follower dichotomy and differ from
 classifications based on regular/structural equivalence.  We then
 show that the interest communities fall into distinct informational
 organigrams characterised by a different mix of user roles reflecting
 the quality of dialogue within them.  Our generic framework can be
 used to provide insight into how flows are generated, distributed,
 preserved and consumed in directed networks.
\end{abstract}


\maketitle

\section{Introduction}

The increasing availability of large-scale relational datasets in a
variety of fields has led to the widespread analysis of complex
networks.  In particular, the current interest in quantitative social
sciences has been fuelled by the importance of social networks and by
the wealth of socio-economic datasets widely available
today~\cite{Brossard2013,Davidian2012,Garcia-Herranz2012,Giles2012,Goel2010,Huberman2012,King2013,Zhou2010,Porter2009}.
Due to the sheer complexity of these networks, it has become crucial to
develop tools for network analysis that can increase our insight into
such data.  A key direction in this area is that of \textit{community detection}, which
aims at extracting a simplified, yet meaningful, coarse-grained
representation of a complex network in terms of `communities' of nodes
at different levels of resolution~\cite{Fortunato2010}.

A common characteristic of many social, engineering and biological networks
is the importance of directionality.  Clearly, it is not the same to
`follow' a widely known personality in Twitter than to be followed by
one. Directionality is also key in food-webs~\cite{Stouffer2012},
brain networks~\cite{Bock2011}, economics
datasets~\cite{Schweitzer2009}, protein interaction
networks~\cite{Schweitzer2009}, and trade networks~\cite{Kaluza2010},
to name but a few. Failure to consider directionality when present
in the data, as is commonly done in numerous network analyses, entails
ignoring the true nature of the asymmetric relationships and
information propagation.  From a methodological perspective, however,
the analysis of directed networks presents unique challenges that put
them beyond standard methodologies. In particular, it is difficult to
extend the structural notion of community (i.e., a group of nodes with
strong connectivity within and with weaker connectivity to the
outside) to the case of directed networks.

Here we show how the analysis of flow patterns on a network can be
integrated to provide a framework for
community~\cite{Delvenne2010,Lambiotte2008} and role~\cite{Cooper2010}
detection. This framework is naturally applicable to directed
networks where flow is an intrinsic feature of the system they
represent.  Our analysis is able to reveal a layered view of the
data from four complementary perspectives: interest communities of
nodes at different levels of resolution; a personalised view of
interest in the network from any vantage point; the identification of
user roles in the network based on directed flows; and a
classification of the interest communities into distinctive
information organigrams.  Our framework is applicable to generic
directed networks, but we showcase our approach through the analysis
of the Twitter network of influential Twitter users during the 2011
riots in England, compiled from the list published by the British
newspaper The Guardian.

\subsection{The directed network of influential
Twitter users during the UK riots}

The riots of August 6-10 2011 in England were followed by an intense
public debate about the role and influence of social media during the
unrest. Politicians, journalists, pundits and bloggers alike weighed
in on the issue but few arguments were based on
data~\cite{Tonkin2012}. A few months after the riots, The Guardian
made available to the public a list of the 1000 `most influential'
(i.e., most {\it re-tweeted}) Twitter users during the
riots~\cite{Evans2011}. The Guardian's list comprised a diverse set of
Twitter users including newspapers, broadcasting services, news
agencies, as well as individual accounts of journalists, politicians,
entertainers, global and local activists, and members of the public.

To enable a quantitative analysis of The Guardian's list, we mined
Twitter in February 2012 and recovered the {\it directed} network
of followers within the list (see SI).  Henceforth we study the
largest connected component of this network consisting of $N=914$ nodes
(Twitter users). The remaining 86 users were either disconnected (i.e.,
they did not follow nor were followed by anyone on the list) or their
accounts had been since deleted.  In our network, an edge indicates
that the source node is subscribed to the {\it tweets} of the target
node, i.e., the direction of the edge indicates the declared interest
whereas information and content travel in the opposite direction
(Fig.~S1 in the SI).

\begin{figure}[tp]
  \begin{center}
    \includegraphics[width=0.8\textwidth]{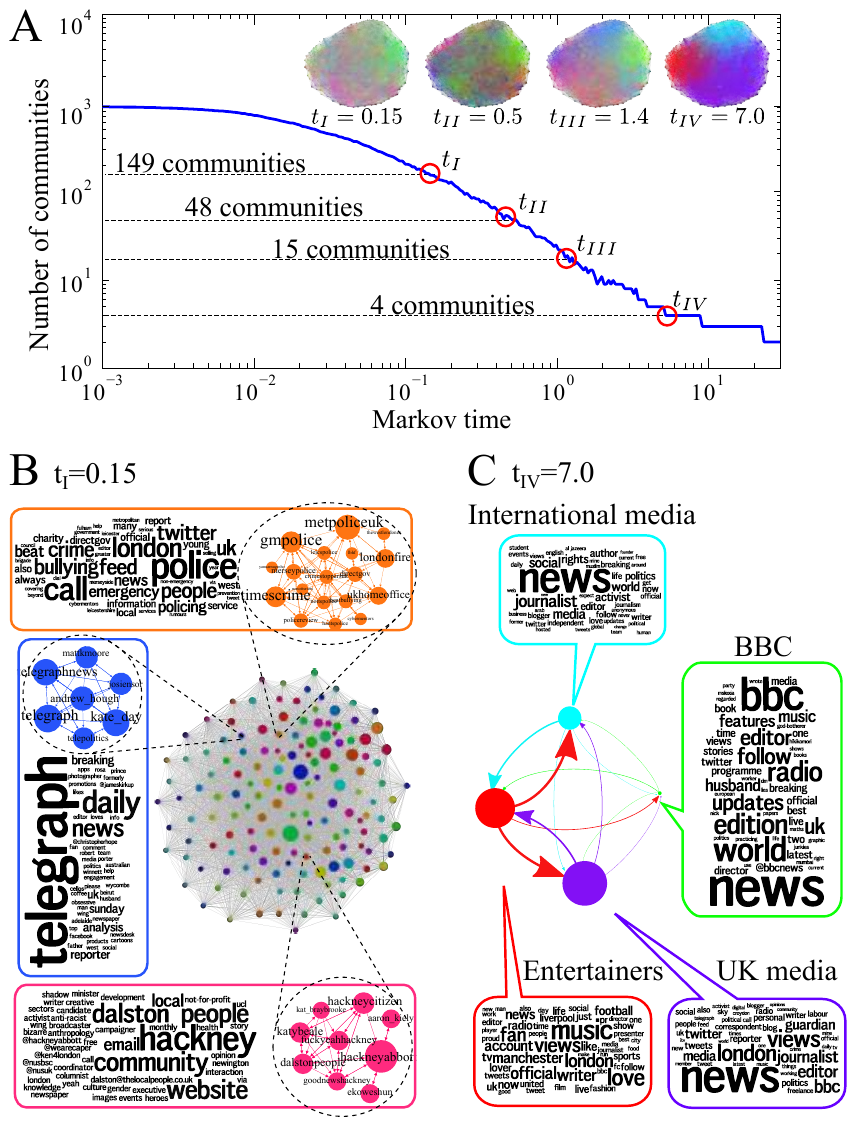}
  \end{center}
\caption{{\small Interest communities at all scales as detected by
  Markov Stability.  {\bf A}: The number of communities at each Markov
  time ($t$).  The inset shows the network with nodes and edges
  coloured according to their community at four
  illustrative Markov times.  Two of these
  partitions at different resolutions are shown in more detail: {\bf
    B.} At relatively short Markov times ($t_I=0.15$) we find 149
  communities (coarse-grained network view in the centre). Three
  examples of communities in this partition are `Police and Crime
  Reporting' (top), `Hackney' (bottom), `The Daily Telegraph' (left)
  shown with their members and their self-description word clouds;
  {\bf C.} At longer Markov times ($t_{IV}=7$) we find 4 communities
  (coarse-grained view in the centre): three large communities broadly
  corresponding to `UK' (bottom-right), `International' (top),
  `celebrities/entertainment' (bottom-left), and a small one corresponding to
  the `BBC' (right). A detailed view of the partitions can be found in the SI.}}
\label{fig:communities}
\end{figure}

\section{Results}
\subsection{Flow-based `interest communities': a view of the network at different resolutions}

To gain a structured view of the communities in the network at
different levels of resolution, we use Markov Stability community
detection~\cite{Delvenne2010,Schaub2012} which has been extended to
deal with directed networks (see Methods, SI and
Ref.~\cite{Lambiotte2008}).  A key advantage of Markov Stability is
that it is based on a quantitative criterion that relies on flow
propagation and containment, and thus identifies {\it flow
  communities}.  The communities so found correspond to `interest
communities', insomuch as information, interest and influence are
propagated, retained and reinforced within them following the
edges. If edge directionality is ignored, the community structure is
blurred and the analysis severely hindered, as shown below.  A second
advantage of our method is that the network is scanned for structure
at all scales, and flow communities are found to be relevant at
different levels of resolution.  Figure~\ref{fig:communities}A illustrates how,
as the network is swept by a continuous-time diffusion process, the
method goes from detecting many small, granular communities (at short
Markov times) to fewer and coarser communities (at longer Markov
times).  As a visual aid to interpret the theme of the communities, we
create `word clouds' from the most frequently-used words in the
Twitter self-biographies of the users in each community. It is
important to remark that the biographies are not used in the network
analysis, i.e., the word clouds serve as an independent, \textit{a
  posteriori} annotation or `self-description' of the communities
found (see SI).

An example of a highly granular partition (149 communities) at short
Markov times is shown in Fig.~\ref{fig:communities}B (and Figs.~S3 and
S4 in the SI).  At this resolution, some communities are defined by
the geographic origin of the Twitter accounts (e.g., Midlands,
Manchester, Liverpool, even Croydon and Hackney within London); others
are determined by employer or institution (e.g., media such as The
Independent, ITV, Channel~4, or The Daily Telegraph); while others
correspond to interest groups (e.g., a community grouping together
police forces and fire departments of riot areas with crime reporters
and civil organisations highlights the police's use of Twitter during
the riots~\cite{Denef2013}).

As the Markov time increases, we find coarser partitions with larger
communities. At $t=0.5$ we find 48 communities, including a
football/sports community (clubs, athletes, sports journalists and
supporters), a politics/Labour community, and a relatively small BBC
community (Fig.~S5 in the SI).  At a longer Markov time
($t=1.3$), we find a partition into 15 communities including the BBC
community, a Sky community, a community of Guardian journalists, a
community of international and alternative media/journalists/activists
(including Wikileaks, Al Jazeera, and Anonymous-related accounts),
among other topical communities (see Fig.~\ref{fig:organigram}).

At even longer Markov times, we show in Fig.~\ref{fig:communities}C a
coarse partition with four communities corresponding broadly to UK
media/activism, international media/activism, entertainment/sports,
and the BBC, which remains as a distinct community across a large span
of Markov times.  We provide a Supplemental Spreadsheet with all
partitions of the network at all Markov times so that interested
parties can explore the all-scale structure of interest communities in
the network.  Furthermore, we have carried a similar analysis using
the well-known information-theoretic Infomap community detection
algorithm ~\cite{Rosvall2008, Rosvall2011}, which in this case leads
to an overpartitioned description with non-optimal compression (i.e.,
a large compression gap) {\it and} unbalanced partitions (see SI for a
discussion)~\cite{Schaub2012,Schaub2012a}.

\subsection{The importance of directionality in detecting interest communities}

In systems that are represented as directed networks, such as Twitter, 
the directionality of the edges is central to their function. 
The full consideration of edge directionality, which
is naturally incorporated in our analysis, is crucial for the
community structure detected.  To illustrate this phenomenon, we compare the
communities found in the original, directed Twitter network with those
obtained when edge orientation is ignored.  We have analysed both
versions of the network (directed and undirected) using the extended
Markov Stability method which can deal with both types of networks.
See the SI (Fig.~S6) for a discussion of the differences in community structure  between the directed and undirected versions of this Twitter network. 
Importantly, relevant communities can go undetected
if one uses standard approaches for community detection based on
undirected structural notions (typically density of
connections~\cite{Schaub2012a}).

\begin{figure}[t]
  \begin{center}
\includegraphics[width=0.8\textwidth]{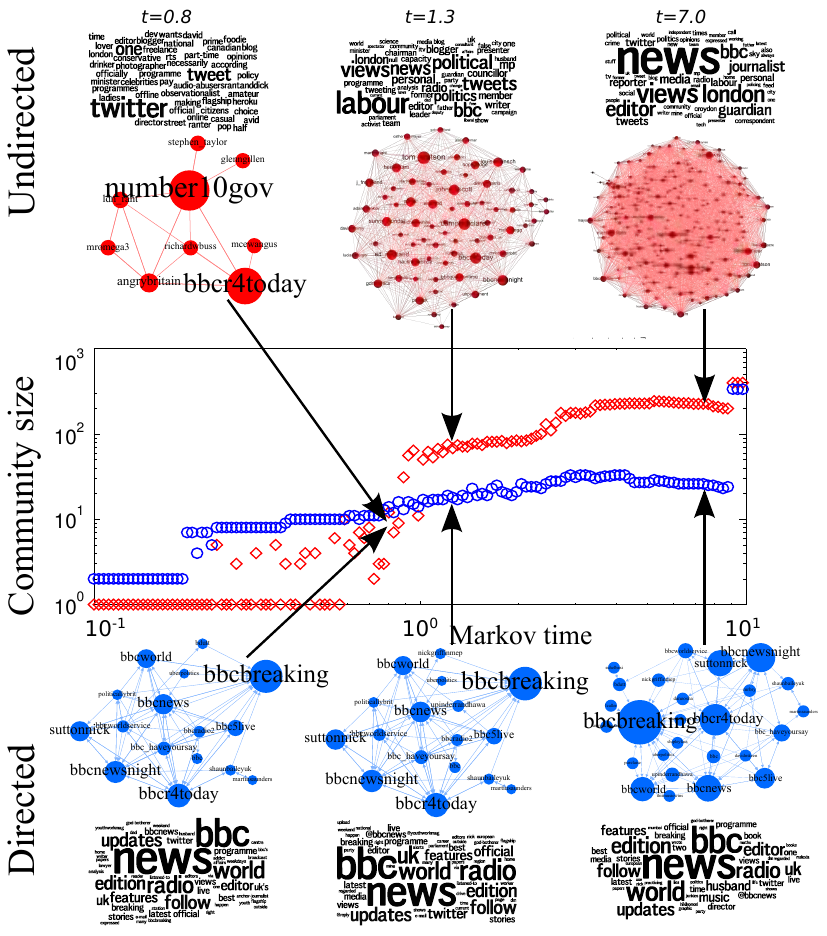}
  \end{center}
     \caption{{\small Communities containing the account of BBC Radio 4's
       Today programme (bbcr4today) in the undirected (top, diamonds)
       and directed (bottom, circles) versions of the network at
       Markov times $t=0.86$, $t=1.3$, and $t=7.0$, along with their
       word clouds.  In the middle we show the size of the communities
       of the Today show in both versions of the network for Markov
       times between $10^{-1}$ and $10^1$. }}
\label{fig:bbcr4today}
\end{figure}

As stated above, the BBC is an example of a flow community that stands
out in its persistency.  In Fig.~\ref{fig:bbcr4today}, we show how the
community of BBC's Today show (a morning news broadcast with a broad
audience) remains consistently grouped across many levels of
resolution in the analysis of the directed network: from an early
Markov time, BBC-related accounts are grouped together and remain so
all the way up to the top levels of resolution, with consistent word
clouds throughout. This phenomenon depends strongly on the
directionality of the flows: the nodes in the BBC community are among
the most important in the network (high in-degree and PageRank),
attracting flow (attention) from elsewhere in the network and
retaining it for long periods of Markov time. In a symmetrised 
network, such communities can go undetected, as shown in
Fig.~\ref{fig:bbcr4today}, where the corresponding undirected
community of the BBC's Today show is quickly blurred across Markov
times and gets mixed with a variety of users with little in common,
consisting mainly of politicians from the Labour Party and
journalists. 

\begin{figure}[tp]
  \begin{center}
    \includegraphics[width=0.8\textwidth]{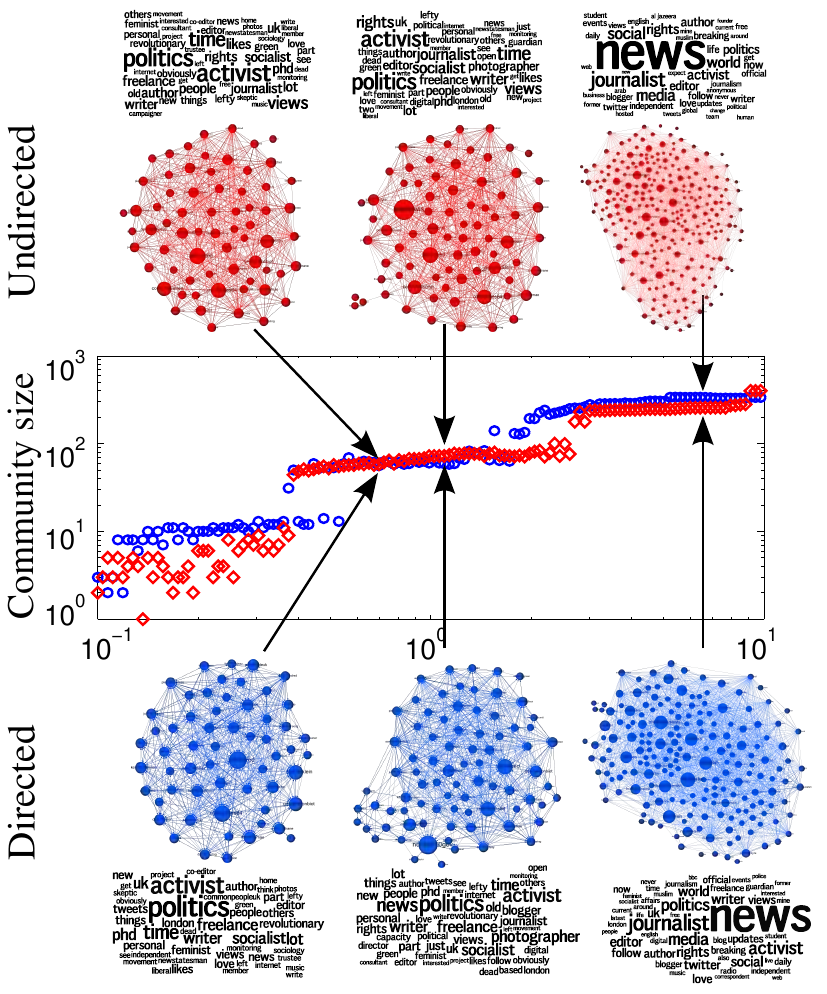}
  \end{center}
\caption{{\small Directed and undirected communities containing the account of
  George Monbiot (georgemonbiot) obtained from the undirected (top,
  diamonds) and directed (bottom, circles) networks at Markov
  times $t=0.8603$, $t=1.3049$, and $t=7.0$, along with their word
  clouds.  Compare these results to those obtained in
  Fig.~\ref{fig:bbcr4today} BBC Radio 4's Today show.}}
\label{fig:georgemonbiot_size_comparison}
\end{figure}

Interestingly, this drastic difference between directed and undirected
communities is not observed for all communities in the network. There
are communities, such as the one including Guardian columnist George
Monbiot, which behave in an essentially similar fashion in
both cases across levels of resolution (see 
Fig.~\ref{fig:georgemonbiot_size_comparison}).
This difference between communities that are sensitive or insensitive
to directionality persists across time scales, signalling the fact
that some groupings (such as the BBC community) are fundamentally
based on retention of directed flows, while others (such as the
Monbiot community) follow from a balanced flow and, thus, can be
captured by standard undirected measures.  We note that the directed
Markov Stabilty method is able to detect both types of communities
simultaneously.

\subsection{Interest distance between nodes: the 
view of the network from a vantage point}

\begin{figure}[tp]
  \begin{center}
    \includegraphics[width=0.8\textwidth]{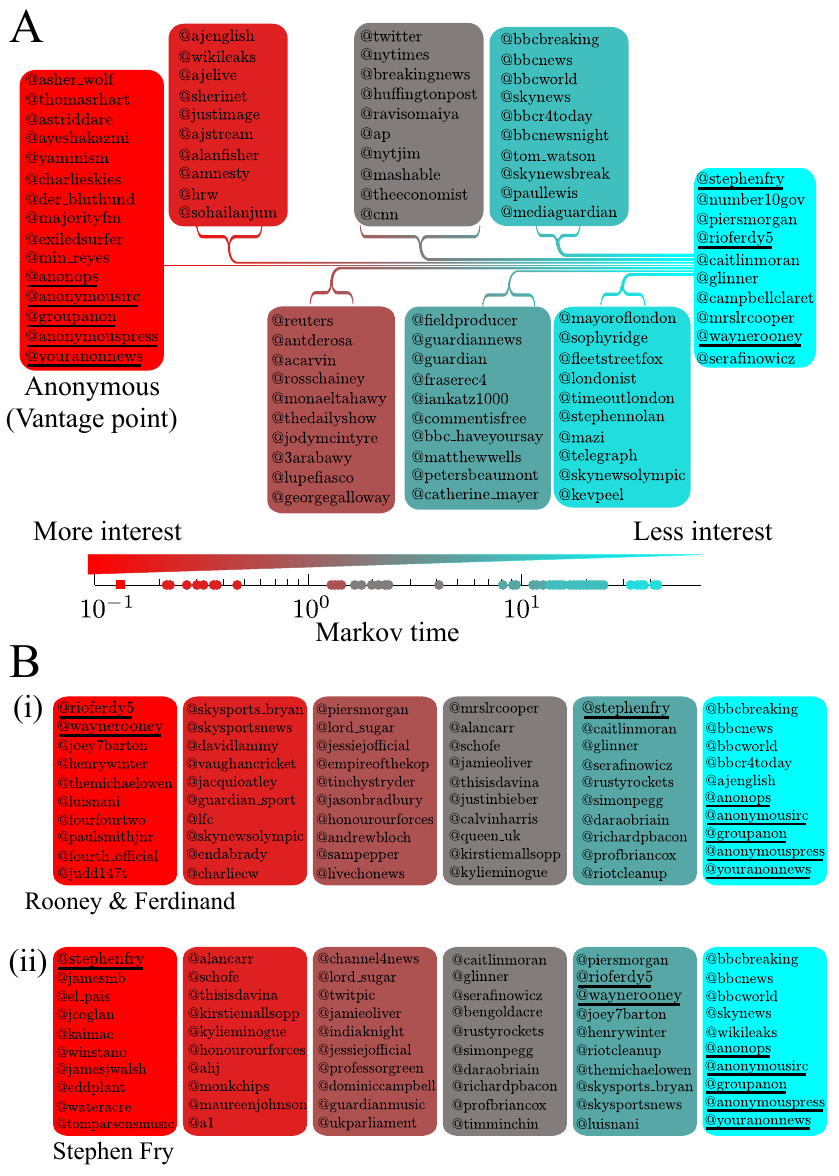}
  \end{center}
\caption{{\small {\bf A}: Personalised view of the network from the vantage
  point of `Anonymous' based on interest distance. The interest
  distance (gradient from red to blue, or dark to light in black and white) 
  is defined as the earliest
  Markov time at which a node belongs in the same interest community
  as `Anonymous'. The number of users in the interest community of
  `Anonymous' (represented by the width of the line) grows as the
  Markov time increases, as users join the community at different
  times.  We show the top ten users (according to PageRank) of every
  batch that joins the Anonymous community.  {\bf B}: The
  reverse personalised views from two vantage points that are of least
  interest to `Anonymous': (i) from the vantage point corresponding to Wayne
  Rooney and several footballers and {(ii)} from the vantage
  point of actor Stephen Fry.  }}
\label{fig:interest_distance}
\end{figure}

As the Markov diffusion explores the network, it can provides us with
information of how interesting the members of the network are to a
given node or groups of nodes (denoted the `vantage point').  Using
our flow-based communities, we establish the \textit{interest
  distance} from the vantage point to any node in the network as the
earliest Markov time at which the node belongs to the same community
as the vantage point (i.e., we compute how `near' they are in an
ultrametric space~\cite{Carlsson2013}).  In
Fig.~\ref{fig:interest_distance}A, we show the computed interest
distance from the vantage point of the Anonymous community (from
$t=0.15$ onwards). Consistent with other
studies~\cite{Coleman2011,Milan2012}, the closest nodes to Anonymous
include Wikileaks, Human Rights Watch, Al~Jazeera and Amnesty
International, followed by a mix of activists and writers, mainstream
international media, and the UK media. Of least interest to Anonymous
are celebrities, UK politicians and footballers.

Unsurprisingly, the picture is starkly different from the vantage
point of the nodes that are of least interest to
Anonymous. Figure~\ref{fig:interest_distance}B shows the interest
distance from the vantage point of footballer Wayne Rooney (of little
interest to Anonymous) whose neighbourhood of interest is dominated by
football, sports and TV celebrities, with news and activists as
distant interests.  The computed interest distance is able to capture
the nuanced information provided by all the directed paths in the
network. This is shown by the fact that Stephen Fry (English actor, TV
personality and writer) is distant from {\it both} Wayne Rooney and
Anonymous (Fig.~\ref{fig:interest_distance}B), while Rio Ferdinand
(Rooney's ex-teammate at Manchester United) is always close to Rooney.
These examples highlight the sensitivity of our Markov exploration and
how the use of vantage points can be used to provide personalised
information about the system.

\subsection{Finding flow-based roles of nodes in directed networks}

A flow-based analysis of directed networks also provides a different
angle for the classification of nodes according to their role in generating and
disseminating information.  Conceptually, it is clear that an account
with millions of followers, such as BBC News, acts as a source of
information (i.e., a reference) while a personal account with only a
handful of followers yet with subscriptions to media outlets acts
mostly as a sink of information (i.e., a listener).  To go beyond this
source/sink dichotomy, or the traditional leader/follower and
hub/authority~\cite{Kleinberg1999} categories, we use here the full
structure of flows in the network to develop a quantitative
methodology that reveals `flow roles' in the network
without imposing the number of roles \textit{a priori}.  Our algorithm
starts by building the {\it role-based similarity} (RBS) matrix (see
Methods below)~\cite{Cooper2010, Cooper2010a}. A feature vector for
each node $i$ is constructed from the scaled pattern of incoming and
outgoing paths of {\it all lengths} and the pairwise cosine similarities  ($y_{i,j}\in [0,1]$)
between all such vectors (see Methods) are stored in the $N\times N$ similarity matrix 
$Y$.  Nodes with similar profiles of incoming and outgoing flows of all lengths 
are classified as having similar \textit{flow roles} in the
network (i.e., when $y_{i,j}$ is close to 1).  The extreme cases
correspond to the standard 'sources' and 'sinks', but an assortment of
nuanced roles spanning these two extremes emerges in our results.  This
analysis provides a complementary use of flows to infer different
properties of nodes: instead of grouping nodes according to flow
persistence (as in the detection of interest communities described
above), RBS provides a grouping of nodes according to their function in
information propagation.

\begin{figure}[tp]
\begin{center}
  \includegraphics[width=0.8\textwidth]{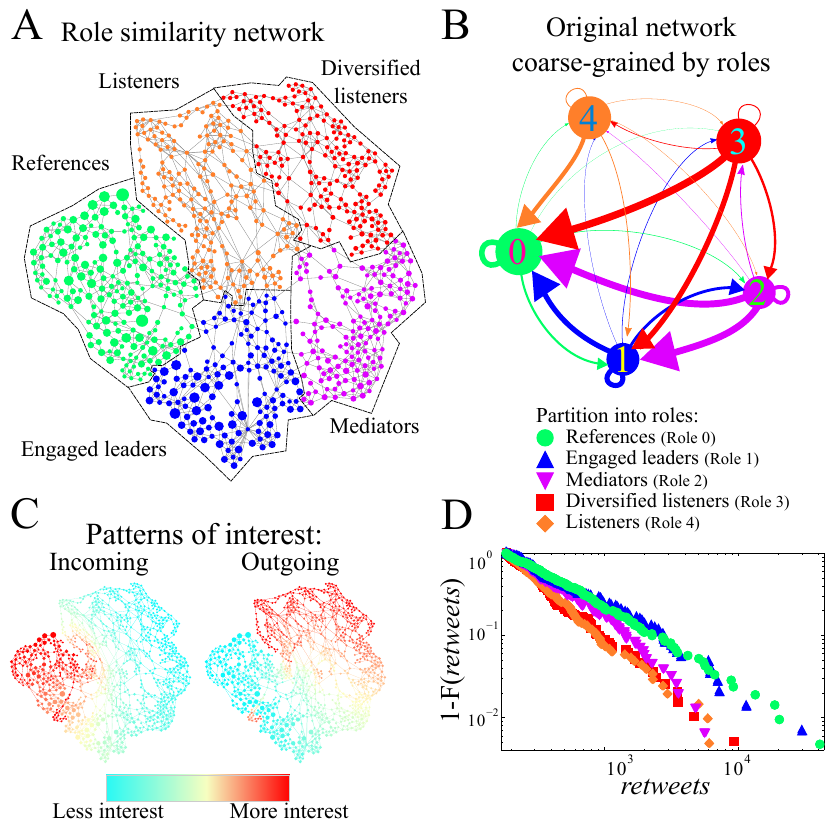}
\end{center}
\caption{{\small Flow-based roles in the Twitter network. {\bf A}: Role
  similarity graph obtained from the path similarity matrix using the
  relaxed minimum spanning tree algorithm.  The size of the nodes is
  proportional to the in-degree in the Twitter network.  Nodes with
  similar profiles of in- and out-paths of all lengths in the original
  Twitter network are close in this role similarity graph.  The role
  similarity graph is found to contain five robust clusters,
  corresponding to flow roles (see Methods and SI).  {\bf B}: The
  original Twitter network coarse-grained according to roles, with
  arrows proportional to users in one role class who follow users in
  another role class.  {\bf C}: Pattern of incoming and outgoing
  interest at all path lengths: (left) nodes in red (dark) 
  receive the most
  attention with higher numbers of incoming paths, while nodes in blue
  (light)
  receive the least amount of attention; (right) nodes in red pay the
  most attention with higher numbers of outgoing paths, while nodes in
  blue pay the least amount of attention with few outgoing paths.
  {\bf D}: Cumulative distribution of retweets for each of the five
  roles: highly retweeted nodes are heavily present in the references
  and engaged leader categories (longer tails) and mostly absent in
  both listener categories. The mediator category lies in-between.}}
\label{fig:riot1000-RBS-t97}
\end{figure}

We have extended the RBS method by using the Relaxed Minimum Spanning
Tree (RMST) algorithm to extract a {\it role similarity graph} from the matrix
$Y$ (Fig.~\ref{fig:riot1000-RBS-t97}A). This novel algorithm creates a 
new graph by emphasising strong similarities between nodes and downplaying 
weaker, redundant similarities based on local continuity and global geometric properties 
of the data similarity $Y$ (see Methods).  
Note that in this RMST-RBS role similarity graph (which is
generated from the Twitter graph but is distinct from it), nodes with
similar connectivity patterns lie close to each other regardless of
how close they are (in a geodesic way) in the original network.  We
then apply graph-theoretical community detection algorithms (such
as Markov Stability) to the RMST-RBS graph and, in doing so, we
reveal groups of nodes (the communities in the role similarity graph)
with similar in- and out-flow patterns corresponding to {\it
  flow-based roles}.  The number of communities on the role similarity graph
correspond to the number of roles in the network. Note that this procedure does not
impose an \textit{a priori} number of roles to be detected (see SI).

Our RMST-RBS analysis finds that there are five flow-based roles for
the nodes in this Twitter network.  Examination
of their incoming and outgoing flow patterns reveals that some of the
groups identified correspond to traditional roles such as {\it
  listeners} (`followers') or {\it references} (`leaders'), but also
distinguishes between different types of leaders, followers and
intermediate roles (Fig.~\ref{fig:riot1000-RBS-t97}A-B). The description of the 
five flow role categories we obtained is as follows:
\begin{enumerate}
  \item[] {\bf References:} Typically, institutional accounts, important
    sources of content, or well-known personalities with many
    followers who follow few accounts, e.g., BBC Breaking News, Al
    Jazeera, Stephen Fry, or the New York Times.

  \item[] {\bf Engaged leaders:} Accounts with large number of followers
    who, unlike references, also follow other users. This category
    includes institutional and personal accounts often meant to
    interact with the public, e.g., Sky News, the British Prime
    Minister's office, Tom Watson (a British MP), or Paul Lewis
    (Guardian editor).

  \item[] {\bf Mediators:} Users who interact with the two leader
    categories (i.e., they follow and are followed by high profile
    accounts), as well as with nodes in the listener categories
    below. Many such accounts belong to journalists and reporters.  
    Examples of mediators include Ross Chainey (Reuters employee), 
    BBC-have-your-say and the London Fire Brigade.

  \item[] {\bf Diversified listeners:} Accounts with few followers that
    follow many nodes from all categories, suggesting diversity in
    their interests and sources of information.

 \item[] {\bf Listeners:} Accounts with few followers (within this
   network, not necessarily over the whole of Twitter) who follow
   mostly Reference nodes. Within this particular network, they can be
   considered as passive recipients of mainstream content.
\end{enumerate}

In the Supplemental Spreadsheet we give the roles of all nodes in the
network. We remark that this classification of nodes into roles is
pertinent {\it only} in the context of the network within The
Guardian's list; it is possible that the role of certain users will be
different if considered embedded in the wider Twitter network, since
the pattern of paths of different lengths attached to each node is
likely to change.

Figure~\ref{fig:riot1000-RBS-t97}C illustrates the mathematical basis for the
classification of nodes into roles by our RMST-RBS algorithm: the
patterns of incoming and outgoing flow at all path lengths are
combined to reveal the different flow roles.  Because RMST-RBS takes
into account the whole spectrum of short to long paths (from length 1
to $K_{max}=133$ in this case, and everything in between) our
classification goes beyond similarity scores that only use single features, such as
in- and out-degrees of the nodes (which appear as the paths of length 1 
in columns 1 and $K_{max}+1$ of the matrix $X(\alpha)$ in
equation~(\ref{eq:RBS})) or eigencentrality-type stationary flow metrics
(columns $K_{max}$ and $2K_{max}$).  Therefore our method obtains information
which is not apparent if we just rank the nodes according to in/out
degree or centrality and then split them into groups.  For example,
ranking the nodes according to PageRank is not enough to distinguish
the `Reference' and `Engaged leader' categories, or to separate
'Mediator' from 'Engaged leader' or 'Diversified listener' (see Fig. S8 in the SI and
Supplemental Spreadsheet).  To confirm the relevance of our findings,
we examine the cumulative distribution of retweets attained by each
node class (Fig.~\ref{fig:riot1000-RBS-t97}D), where we see a clear
separation between the leader (reference and engaged leader nodes) and
follower (diversified listeners and listeners) categories, with the
mediators lying in between both groups. It is important to remark that
the retweet data in Fig.~\ref{fig:riot1000-RBS-t97}D is not part of
our role detection and is only used \textit{a posteriori} to inform
our understanding of the flow roles obtained (see also Fig.~S8 in the
SI).

The flow roles we find here are conceptually and practically different
from those obtained using well established theories in social network
analysis, e.g. Structural Equivalence (SE)~\cite{Lorrain1971} and
Regular Equivalence (RE)~\cite{Everett1994, Everett1996, Luczkovich2003, Borgatti1993}.  SE bases node similarity
on overlapping sets of neighbours (i.e., two nodes are similar if many
of their neighbours are the same), while RE-based methods rely on node
colorations and neighbourhoods (i.e., two nodes have the same role if
the colours of their neighbours are the same, regardless of the number
of common neighbours).  Hence SE and RE are essentially short-path
methods and not suitable for networks like the one studied here where
the full structure of flow is essential (see SI for a detailed
description of RE and SE roles and their lack of information content
in this network). Furthermore, RE methods are not robust to small
random perturbations in network connectivity due to their
combinatorial nature.

\begin{figure}[tp]
  \begin{center}
    \includegraphics[width=0.8\textwidth]{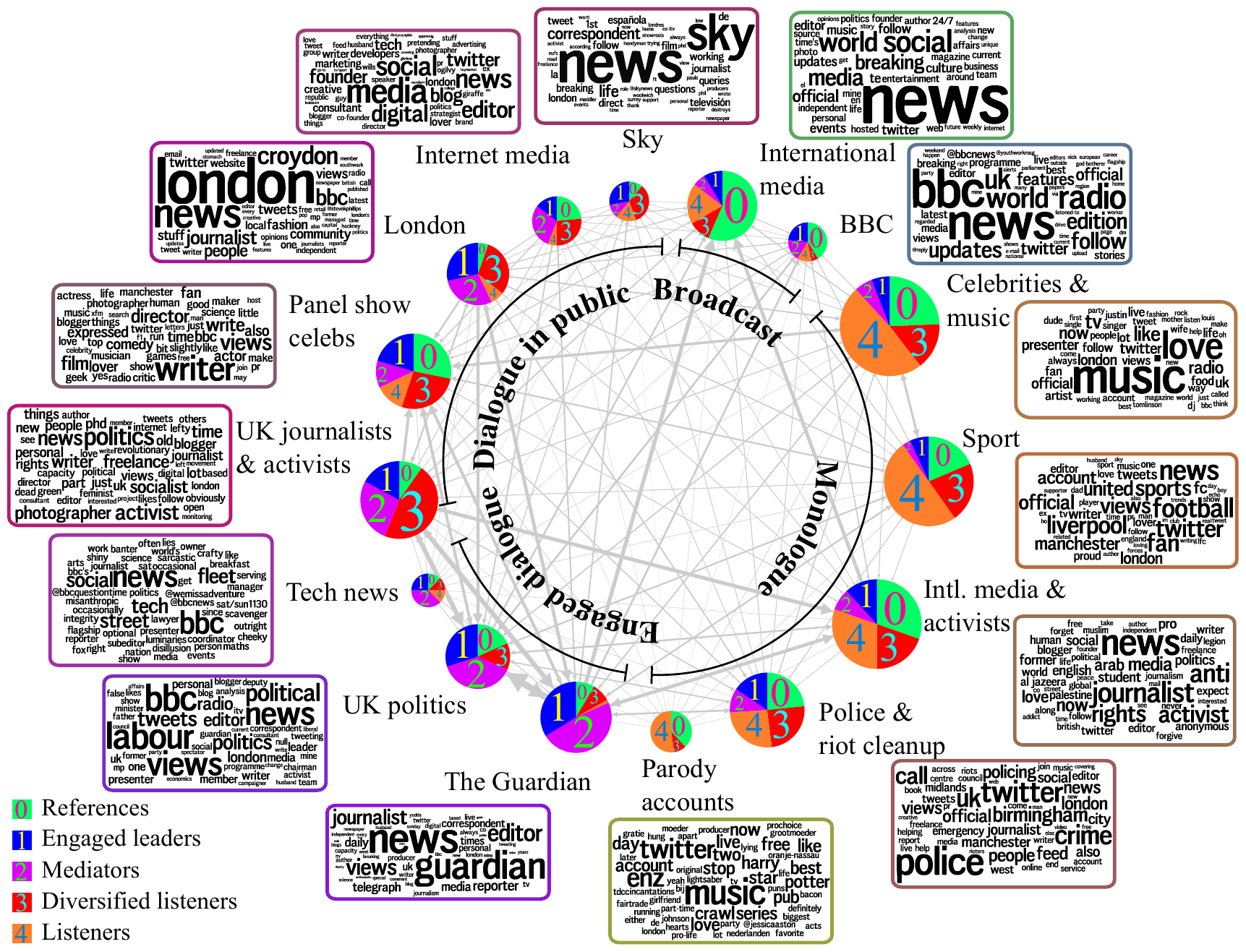}
  \end{center}
\caption{{\small Mix of roles of the 15 interest communities found at
    $t=1.3$.  The communities reflect a diverse set of topical
    groupings (see word clouds with the top 50 nontrivial words in the
    user biographies) and are characterised by different mixes of the
    five flow roles, as shown by the pie charts. The organigrams range
    from reference-listener schemes (`broadcast' and `monologue') to
    more balanced dialogue communities (`engaged dialogue' and
    `dialogue in public') in which engaged leaders, mediators and
    diversified listeners dominate. }}
\label{fig:organigram}
\end{figure}

\subsection{Interest communities and their distinct mix of roles}

Heretofore, our two-pronged flow-based analysis has led to groupings of the nodes
according to two criteria: interest communities (at different
resolutions) and flow roles .  Both perspectives
present complementary views of the information in the network and can
be combined to characterise the internal organisation of interest
communities in terms of the mix of roles of their members.
Figure~\ref{fig:organigram} presents this integrated view for the 15
interest communities at medium resolution (Markov time $t=1.3$), and
the five node roles found through RBS-RMST. Using a simple $k$-means
clustering of their role-mixes, we find that the 15
communities fall into four types of informational organigrams (see SI). 
Two of these organigrams broadly conform to communities
formed mostly by leaders (`references') and their followers
(`listeners'), though with some important differences: {\it
  ``monologue communities''} are predominantly composed of references
with a set of loyal (non diversified) listeners with information
flowing mostly in one direction (e.g., `Celebrities/Entertainers',
`Sport', 'Parody accounts'), while in {\it ``broadcast communities''}
most members are references delivering content broadly to a wide
variety of users in the network (e.g `BBC' and `International media').
In addition, there exist two organigrams with a more balanced dialogue
structure: {\it ``dialogue in public''}, which involves many
diversified listeners (e.g., `Panel show celebrities', `London', or
groups heavily based on Internet interaction such as `UK journalists
\& activists'); and {\it ``engaged dialogue''}, which is dominated by
engaged leaders and mediators (e.g., `Politics' and `The Guardian').
These two dialogue organigrams reflect the importance of online
interaction in information networks, where bottom-up grassroots
associations, bloggers and commentators from the public interact
directly with accounts linked to news outlets and official political
organisations.

\section{Discussion}

In this work, we have used the Twitter network constructed from the
list of influential users during the UK riots of 2011 collected by The
Guardian to showcase how flow-based methods in directed networks can
lead to enhanced insight into the structure of data.  Our analysis
reveals interest communities into which users fall at different levels
of resolution. The interest communities found confirm the relevance of
news organisations and media, yet providing a layered view in terms of
their focus (UK/international, mainstream/alternative) and of
relationships to each other and to the overall network. The enhanced
sensitivity of our multi-resolution analysis allows us to uncover
small but significant groups related to local organisations or
clean-up groups in riot areas which appear close to police and law
enforcement groupings. In addition, our analysis reveals groupings
that have an unexpected relevance in a network that was selected on the basis
of `retweeting' importance during an event of civil unrest. In
particular, a significant grouping of celebrities, sport personalities
and pop musicians act as the center of a significant interest community.  Also
intriguing is the role of interest groups based on humour in such
situations, as represented by communities of comedians and parody accounts. 
Our work points at future studies on how to use this
type of analyses to improve and tailor communication strategies during
times of unrest, in particular with regards to providing a
personalised view of the network from any given vantage point (i.e.,
from any node or group of nodes) based on the interest distance of information flow. 
These results can be a
starting point to examine textual information and analyse the influence of groups of interest on
observed behaviors in this and similar datasets.

Using flow transfer in the network, we find that the Twitter users in
this network fall into a palette of five flow roles, whereas interest
communities exhibit distinct mixes of such roles reflecting diverse
communication patterns within them.  Some
communities contain one-way communication patterns (e.g., celebrities
and their followers), whereas other communities harbour more balanced
dialogue patterns.  In particular, our analysis highlights the
differences between media organisations and their distinct patterns of
interaction with the influential users in this network.  For instance,
international mainstream media tend to fall into the Broadcast and
Monologue categories, as would be expected in a network of UK-based
events. On the other hand, the UK and specialised media exhibit a more
diverse pattern of interactions with their followers: some of them are
highly engaged with mediators and diversified listeners whereas others
largely maintain the more traditional role of publishing content.

This work also highlights the use of multiscale network analyses, which
go beyond local information of individual users towards aggregate global
metrics, to deliver an enriched view of information dissemination in
social networks, thus uncovering relationships and roles of nodes and
providing concise coarse-grained descriptions of the network.  We
hope that our results (all available in the Supplemental Material)
could be a helpful resource to aid in the study of online interactions
during the UK riots of 2011.

More generally, our work highlights the importance of directionality
in network analysis.  When the notion of flows (e.g., of people,
information, energy, goods) is central to a network, ignoring
directionality destroys information, `blurring' the structure,
especially at the finer levels of resolution, so that key communities
(e.g., the BBC, Sky, and geographical communities in our analysis)
will go undetected.  The formulation of community and role detection
in terms of flow dynamics thus provides an integrated methodology for
the analysis of systems (natural or man-made) with directed network
representations.


\appendix

\section{Methods}

\subsection{Community detection with Directed Markov Stability}

We give here a summary of the theoretical ideas and computations
underpinning our analysis of interest communities using directed
Markov Stability. For a full explanation of the method, see
Refs.~\cite{Delvenne2010,Lambiotte2008,Schaub2012}.  The code for the
Markov Stability algorithm can be downloaded
from~\cite{StabilityWebsite}.  For an expository article, see
Ref.~\cite{Delvenne2013}.

\subsubsection{Graph theoretical definitions}

Let $A$ be the $N \times N$ adjacency matrix of a directed network
($N=914$ in the riots Twitter network), where $A_{i,j}=1$ if node $i$
has an edge to node $j$ and 0 otherwise.  Note that $A\neq A^T$ in
general.  In a directed network, each node has an in-degree
($k_{in}=A^T\mathbf{1}$, where $\mathbf{1}$ is the $N\times 1$
vector of ones) and an out-degree ($k_{out}=A\mathbf{1}$) which are
the number of edges directed at the node and departing from the node,
respectively.

\subsubsection{Random walks on directed graphs}

A Markov chain on the graph is usually
defined by the transition matrix $M=D^{-1}A$, where $D =
\mathrm{diag}(k_{out})$ is the diagonal matrix of node out-degrees.
For nodes where $k_{out}(i)=0$, the convention is to set $D(i,i)=1$.
The evolution of a discrete-time Markov chain is given by
\begin{align}
\mathbf{p}_{t+1} =  \mathbf{p}_t \, D^{-1}A  = \mathbf{p}_t \,M
\end{align}
and, alternatively, a Markov process in continuous time is governed by
the Kolmogorov equation:
\begin{align}
 \dot{\mathbf{p}} = - \mathbf{p}\;[I_N - D^{-1} A] = -
 \mathbf{p}\;[D^{-1}L].
 \label{eq:kolmo}
\end{align}
Here $\mathbf{p}$ denotes the $1\times N$ dimensional probability
vector, $I_N$ is the $N\times N$ identity matrix, and $L$ is the
(combinatorial) Laplacian matrix of the graph. We can view both these
processes as defining a random walk taking place on the graph.

To ensure that the random walk is ergodic, we add a `teleportation'
component to the dynamics~\cite{Lambiotte2008}
to obtain a new transition matrix 
\begin{equation}
  B = \lambda M + \left[(1-\lambda) I_N + \lambda \,
    \mathrm{diag}(a)\right]\frac{\mathbf{1}\mathbf{1}^T}{N}.
  \label{eq:teleport}
\end{equation}
Here $\lambda \in (0,1)$ is the probability that a random walker
arriving at a node will follow an outgoing edge, while the walker will
be `teleported' (i.e., it will jump to any other node in the network
chosen at random) with probability $(1-\lambda)$.  In this work, we
use $\lambda=0.85$ throughout.  The probability that any node is
visited by a teleported random walker is drawn from a uniform
distribution (i.e., each node has the same probability $1/N$ of being
visited, though other choices are possible~\cite{Lambiotte2012}).
The $N \times 1$ vector $a$ is an indicator for dangling
 nodes: $a(i)=1$ if $k_{out}(i)=0$ and $a(i)=0$ otherwise. Upon
 visiting a dangling node, a random walker will be teleported with
 probability 1.

In this work we consider the continuous time process in 
equation~(\ref{eq:kolmo}) with transition matrix $B$:
$$\dot{\mathbf{p}} = -\mathbf{p} \left(I_N - B\right).$$
The steady-state $\pi$ is given by the leading
left-eigenvector of $B$ (associated by the eigenvalue 1), 
and the time-dependent transition matrix is $P(t) =
\exp\left(-t(I_N-B)\right)$. 

\subsubsection{Directed Markov Stability and community detection}

We have recently introduced the community detection method known as
{\it Markov Stability}. The basic idea is that the study of the
dynamics of diffusion processes on networks can be used to identify
meaningful partitions at different
resolutions~\cite{Delvenne2010,Lambiotte2008}.  This notion can be
illustrated by the example of observing how a droplet of ink would
diffuse in a container. If the container has no structure, the ink
diffuses isotropically.  If the container is compartmentalised, the
dye would not spread isotropically but would rather get transiently
trapped for longer times in certain parts of the container until it
eventually becomes evenly distributed throughout the whole vessel.
Hence the time dynamics of this diffusion process provides us with
valuable information about the structural organisation of the
container.  A similar idea can be applied to the diffusion on a graph.

From this dynamical perspective, the Markov time acts as a means to
scan the structure of the graph \textit{at all scales}, thus providing a
dynamical zooming over the structure of the graph. In this process of
zooming, the diffusion explores increasingly larger sections of the
graph and identifies increasingly coarser partitions.  Communities are
identified as subgroups within which the probability flow is well
mixed yet the flow remains contained over particular time scales. The
communities are found by finding the partitions that optimise a
time-dependent cost function.  As the diffusion progresses, this cost
function optimisation allows us to rank the goodness of partitions and
to identify which partitions are relevant over different time scales.
Relevant partitions appear as robust, because they are optimal over
extended time-intervals and/or in terms of the basin of attraction of
the optimisation process.

A partition of a network into $C$ communities can be encoded into the
$N \times C$ indicator matrix $H$, where $H_{i,c}=1$ if node $i$
belongs to community $c$ and 0 otherwise. Then the Markov Stability of
the partition is defined as the trace of the clustered
autocovariance of the diffusion process taking place on the
 graph~\cite{Delvenne2010}:
\begin{equation}
  r(H, t) = \trace{H^T\left[\Pi P(t)  - \pi^T\pi\right]H},
  \label{eq:stability}
\end{equation}
where $\Pi = \textrm{diag}(\pi)$. 

We find the communities in the
network at all scales by optimizing the Markov
Stability~(\ref{eq:stability}) for any given value of $t$ (the Markov
time) over all partitions $H$. This is an NP-complete combinatorial
problem~\cite{Delvenne2010} and to provide optimised solutions, we use
the Louvain greedy optimisation heuristic~\cite{Blondel2008}, which
works well in practice.
Note that although the original Louvain
method is formulated only for symmetric $\Pi P(t)$, we have shown that
the optimisation of the Markov Stability in the case of directed
networks can be reformulated in terms of the symmetrised matrix $W =
\left(Q + Q^T\right)/2$, where $Q = \Pi
P(t) - \pi\pi^T$, which follows from $\trace{ H^TQH}=
\trace{H^T W H}$ ~\cite{Lambiotte2008}.

The Markov Stability framework explores the community structure of a
network at all scales through the dynamic zooming provided by the
duration of the diffusion process $t$: if $t$ is small, the diffusion
process is short and the optimal partitions consist of many small
communities; for larger values of $t$ the diffusion process explores
the network further and, consequently, we find fewer and larger
communities (see Fig.~\ref{fig:communities} and
SI)~\cite{Delvenne2010,Lambiotte2008}.  The fact that Markov Stability
is based on the analysis of flows diffusing in the network allows us
to extend seamlessly the analysis of communities to directed
networks. In our framework, the defining characteristic of communities
is the persistence of flow (contained and well-mixed) within the
community over a given timescale.  Importantly, because Markov
Stability is based on the concept of flow, it can detect non-cliquish
communities, i.e., communities that are not characterised by density
of links but by retention of flow~\cite{Schaub2012}. As we show in the
Main Text and the SI, this property is vital for the analysis of
networks with flows of information, particularly in the directed case.

As our method scans dynamically through Markov time, it enables us to
find communities defined by their flow patterns at all scales through
the optimisation of the stability $r(H,t)$ for a range of $t$ spanning
several orders of magnitude.  Briefly, for each value of $t$ we find
the partition of the network that maximises~$r(H,t)$ using the Louvain
method~\cite{Blondel2008} from 100 different random initial guesses.
The consistency and robustness of the 100 partitions obtained from the
optimisations is assessed with the normalised variation of information
(VI)~\cite{Meila2007}, as described in
Refs.~\cite{Lambiotte2008,Schaub2012}; see Fig.~S2 in the SI.  The VI
allows us to gauge the consistency of the partitions obtained from
optimising $r(H, t)$ at each $t$. A decrease in VI (or an inflection
point) at a particular value of $t$ suggests relevant community
structure at this time scale.

The computational complexity of Markov Stability in its full form (as
used here) is slightly better than $O(N^3)$ due to the computation of the
matrix exponential. This is appropriate for graphs up to several thousands of nodes. 
For larger graphs, Refs.~\cite{Lambiotte2008, Delvenne2013} discuss an
approximate (linearised) version of Markov Stability which is approximately $O(N)$ and 
can be applied to much larger graphs~\cite{LeMartelot2013}.

\subsection{Finding flow roles in directed networks with RBS-RMST}

To classify the nodes according to roles, we combine role-based
similarity~\cite{Cooper2010, Cooper2010a} with the relaxed minimum
spanning tree algorithm and Markov Stability.  
We start by creating the { role-based similarity} (RBS) matrix, which
exploits the directed structure of the graph to obtain a similarity
score that measures how alike the flow connectivities of nodes are.

For each node, we obtain a profile vector
that contains the number of incoming and outgoing directed
paths (incoming and outgoing) of lengths from 1 up to
$K_{max}<N-1$ for all nodes. The number of paths corresponding to each
node is scaled by a constant and stored as row vectors to create the $N
\times 2K_{max}$ matrix:
{\small
\begin{align}
  X(\alpha) &= \overbrace{\left[ ~\dots~\left(\frac{\alpha}{\lambda_1} A^T
      \right)^K \mathbf{1}~\dots~ \right.}^\text{paths in} \overbrace{\left|
      \left. ~\dots~\left(\frac{\alpha}{\lambda_1} A \right)^K \mathbf{1}~\dots~\right] \right.}^\text{paths out},
  \label{eq:RBS}
\end{align}
}
where $\alpha \in (0, 1)$, and $\lambda_1$ is the largest eigenvalue
of $A$. The
cosine distances between any two rows of $X(\alpha)$ are stored in the
$N\times N$ similarity matrix $Y(\alpha)$:
\begin{equation} y_{i,j} = \frac{\mathbf{x}_i
  \mathbf{x}_j^T}{\norm{\mathbf{x}_i}_2\norm{\mathbf{x}_j}_2}. 
  \label{eq:Y}
\end{equation}
By construction $y_{i,j} \in [0,1]$ with $y_{i,j}\simeq 1$ whenever
nodes $i$ and $j$ have very similar profiles of incoming and outgoing
paths of all lengths, i.e., when nodes $i$ and $j$ play similar roles
in the network in terms of flow generation, distribution and
consumption.  If we choose a small $\alpha$, the terms
$(\frac{\alpha}{\lambda_1} A^T)^K$ converge quickly and the maximum
path length ($K_{max}$) is small.  Hence we would classify nodes
according only to their immediate neighbourhoods (in the limit of
$\alpha \to 0$, nodes are classified according to $k_{in}$ and
$k_{out}$ only).  If, on the other hand, $\alpha$ is close to 1, the
resulting $K_{max}$ is larger, thus including global information of
the network to classify the nodes.  We have followed the iteration prescription
detailed in~\cite{Cooper2010} and 
used $\alpha=0.95$, which gives $K_{max}=133$.  In its current form, t
he computational complexity of RBS in its current form is 
slightly better than $O(K_{max}N^3)$. Further algorithmic improvements
of this method will be the object of upcoming publications.

\subsection{The relaxed minimum spanning tree similarity graph}

The similarity matrix $Y$ is then transformed into a role similarity
graph (Fig.~\ref{fig:riot1000-RBS-t97}) by using the RMST algorithm,
which uses a geometric graph embedding based on the iterative addition
of relevant edges to the backbone of the minimum spanning tree: edges
are only added if there is no alternative path on the tree with a
lower distance. This construction attempts to preserve the continuity
of the dataset, thus unfolding the structure of the data.  The
similarity network thus constructed is then analysed for communities
using Markov Stability. 

In sum, from the original adjacency matrix of the network we use RBS
to compute the pairwise similarity between the nodes in the matrix
$Y$; then we use the RMST to extract the role similarity graph; and on
this graph we perform community detection to obtain the roles of the
nodes.  We find that the similarity graph of the
Twitter network has a robust partition into five types of roles (at
Markov time $t=97$, with zero variation of information, see SI). The role
classification for every node is provided in the Supplemental
Spreadsheet.

The basic idea of RMST is that weak cosine similarities between high
dimensional vectors are non-informative and do not contribute to our
understanding of the structure of the dataset---in high dimensional
space weak similarities are commonplace thus clouding the
relationships in the network.  Our strategy for the role similarity
graph primes the importance of strong similarities: two nodes will not
be linked directly in the role similarity graph, if there is already a
chain of strong similarities (a weighted path) that links them.  More
precisely, consider the distance matrix $Z$, where
$z_{i,j}=1-y_{i,j}\in [0,1]$ is the distance between nodes $i$ and $j$
according to their flow profile vectors.  The classical strategy for
network construction from a distance matrix is to include an edge
between two points if the pairwise distance is less than a threshold
value (e.g., if $z_{i,j} < \varepsilon$). The problem with this crude
strategy is that it does not recover the geometry of the data when the
points are not homogeneously distributed~\cite{Tenenbaum2000}. 
If the threshold is small,
the network will consist of several disconnected components. If the
threshold is large then the network will contain densely connected
components, which would take us back to the same problem we had with
the full matrix. These problems appear because of the local nature of
such an approach, which is exclusively based on local distances.

Instead, we use a global strategy for the construction of the role
similarity graph using the Relaxed Minimum Spanning Tree (RMST)
algorithm, a method well-suited for extracting meaningful networks
from datasets that are not homogeneously distributed in a
high-dimensional space (in this case $\R^{2K_{max}}$).  We begin with
a Minimum Spanning Tree (MST) as the initial backbone of the graph,
and we add edges iteratively using the following simple heuristic
(note that the MST is such that the sum of edges in the tree is
minimal, and a path in the MST is the path between two nodes that
minimises the maximal edge weight).  At each step of the iteration, we
consider whether the MST path between any pair of nodes $i$ and $j$ is
a significantly better model compared to the direct edge $z_{i,j}$. If
the maximal edge weight in the MST path is significantly smaller than
$z_{i,j}$, the MST path is considered a better model based on the
continuity achieved through short distances. If, on the contrary, the
maximal edge weight along the MST path is comparable to $z_{i,j}$,
then we consider that there is not sufficient evidence to say that the
MST path is a better model for data continuity and we add an edge
between $i$ and $j$ in the RMST. Therefore, the edges in the RMST are
generated as:
{\small 
\begin{equation}
RMST_{i,j} = \left\{ 
 \begin{array}{rl}
   1 & \mbox{if $\quad \textrm{mlink}_{i,j} + \gamma(d^{k}_{i} + d^{k}_{j})
     > z_{i,j}$}, \\ 
   0 & \textrm{otherwise,}
   \end{array}
 \right.
 \label{eq:RMST}
\end{equation}
}
where $\textrm{mlink}_{ij}$ is the maximal edge weight in the MST path
between nodes $i$ and $j$, $d^{k}_{i}$ is the distance from node $i$
to its $k$-th nearest neighbour, and $\gamma$ is a positive
constant (here, $k=1$ and $\gamma=0.5$). The factor $\gamma d^{k}_{i}$
approximates the local distribution of data points around every
point. Our approximation of the local distribution around a point is
motivated by the Perturbed Minimum Spanning Tree
algorithm~\cite{Carreira2005}. This iteration is continued until no
more edges are added to the RMST.  
We call this the \textit{role similarity graph}. 
The complexity of the
RMST algorithm is $O(N^2)$. 

\section*{Acknowledgments}

MBD, SNY and MB acknowledge support from the UK EPSRC through grant
EP/I017267/1 under the Mathematics Underpinning the Digital Economy
program. BV was funded by a PhD studentship of the BHF Centre for
Research Excellence.  MBD also acknowledges support from the James
S. McDonnell Foundation Postdoctoral Program in Complexity
Science/Complex Systems-Fellowship Award (\#220020349-CS/PD Fellow).
The authors thank Michael Schaub for many useful conversations.

\setcounter{figure}{0}
\setcounter{section}{0}
\setcounter{equation}{0}
\renewcommand\thesection{S\arabic{section}}
\renewcommand\thefigure{S\arabic{figure}}
\renewcommand\theequation{S\arabic{equation}}

\section*{{\Large Supplementary Information}}

The Supplementary Spreadsheet accompanying this text can be downloaded
from: {\tt http://www2.imperial.ac.uk/\~{}mbegueri/Docs/riotsCommunities.zip}
or {\tt http://rsif.royalsocietypublishing.org/content/11/101/20140940/suppl/DC1}
    
\section{The UK riots Twitter network}

\subsection{The Guardian's list of influential user during the UK riots of 2011}
Recently, the British newspaper The Guardian announced that it
possessed a database of 2.5 million Twitter messages (tweets) related
to the riots that took place in England during the summer of 2011.  In
December 2011, The Guardian made public a list of the 1000 most
influential Twitter users according to the number of forwarded
(re-tweeted) status updates during the riots~\cite{Evans2011}.  This
diverse list of Twitter users includes large news and media outlets,
social and political organisations, as well as personal accounts of
politicians, journalists, activists, celebrities, and parody
accounts. In addition, a considerable fraction of the list is formed
by regular individuals whose tweets received attention during the
riots, and about whom little is known. The list provided by The
Guardian contains the Twitter user name, the number of re-tweets, and
a mini-biography of at most 140 characters which Twitter users can
provide to describe themselves.

\subsection{Creating the Twitter network from The Guardian's list}

\begin{figure}[tp]
  \begin{center}
    \includegraphics[width=0.8\textwidth]{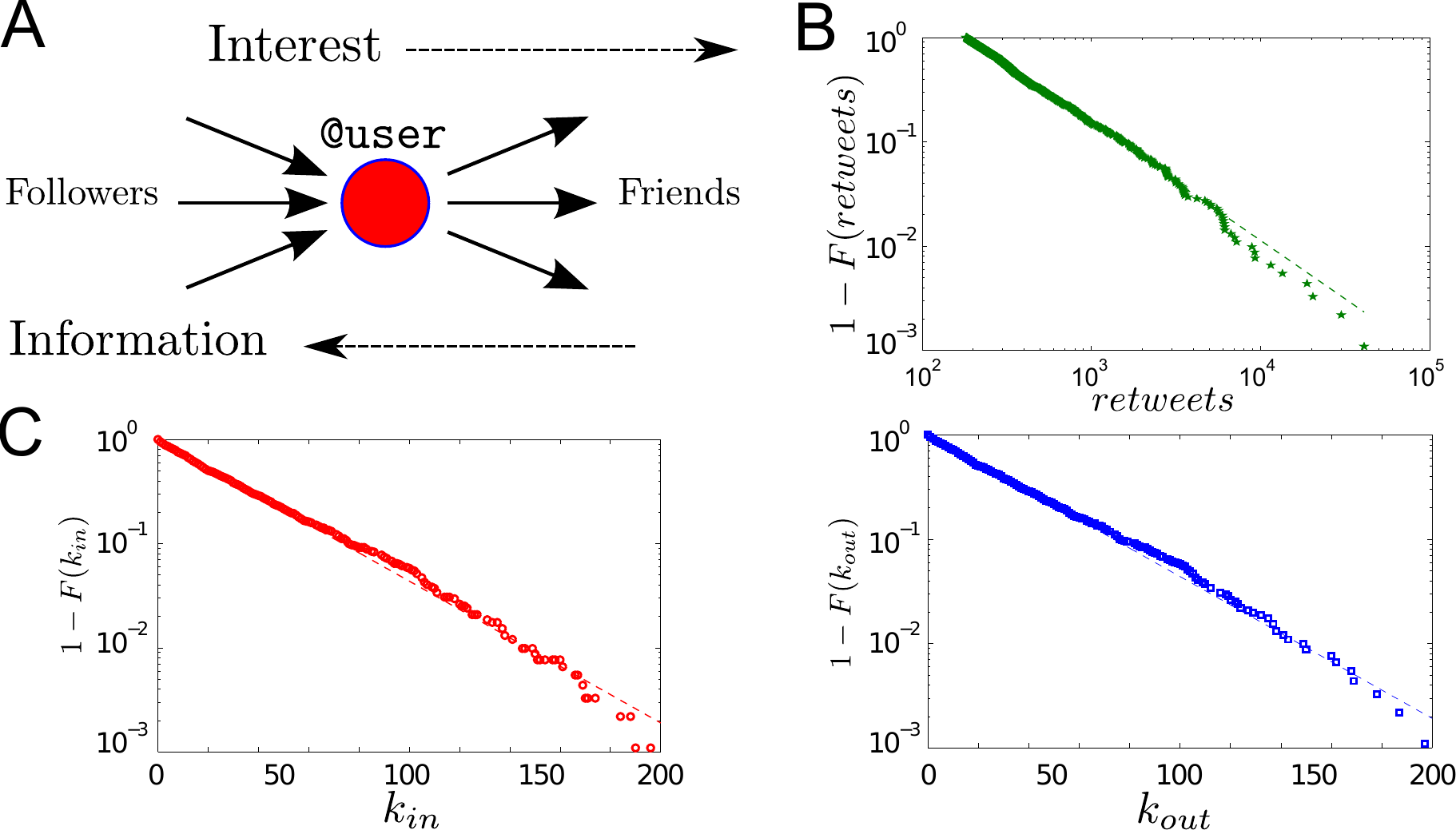}
\end{center}
\caption{Some properties of the UK riots' influential Twitter users
  network.  {\bf A}: A directed network is created by intersecting
  each user's friends with the user list. The interest (or attention)
  is directed at friends, and information travels in the opposite
  direction (from the friend to the follower).  {\bf B}: Cumulative
  re-tweet distribution. The dotted line is the fit
  $retweets^{-\alpha}$, where $\alpha = 2.12$.  {\bf C}: Cumulative
  in-degree (left, in red) and out-degree (right, in blue)
  distributions.  The dashed lines show fits of the data to
  $e^{-\lambda_{in}k_{in}}$ and $e^{-\lambda_{out}k_{out}}$, with
  $\lambda_{in} = 0.0313$ and $\lambda_{out} = 0.0312$.}
\label{fig:distributions}
\end{figure}

We created the network from The Guardian's list (Figure 1 of the Main
Text) by mining Twitter in February 2012 and obtaining the ``friends''
of all users in the list, i.e., the Twitter accounts to which each
user is subscribed, which is publicly available information.  A
directed, unweighted network was obtained by intersecting the list of
friends with the members of the list.  In this network, the
information flows from target to source along each edge, i.e.,
opposite to the direction of the edge, which represents the declared
interest of the source node (Fig.~\ref{fig:distributions}A).  The
resulting directed network has a giant component of 914 nodes. The
remaining 86 nodes are either completely disconnected (i.e., they do
not follow nor are followed by anyone on the list), or their accounts
have since been discontinued.  All the subsequent work in this paper
uses this 914-node network.

\subsection{Some statistics of the network}
The distribution of the number of retweets of the members of the list
(Fig.~\ref{fig:distributions}B) is compatible with a power-law with
exponent $\alpha \sim 2.12$ ($p=0.75$, using the criterion in
Ref.~\cite{Clauset2009}), which is consistent with previous analyses
of Twitter data sets~\cite{Zhou2010}.  The cumulative in- and
out-degree distributions of the connected component are shown in
Fig.~\ref{fig:distributions}C. Although both distributions appear
exponential with, respectively, parameters $\lambda_{in}\approx
0.0313$ and $\lambda_{out}\approx 0.0312$, a Kolmogorov-Smirnov
statistic does not provide statistically significant support for this
hypothesis ($p<0.1$).  The distributions of $k_{in}$ and $k_{out}$ in
this data are less skewed than other published
results~\cite{Kwak2010}.  Note also that the in- and out-degrees are
similarly fat-tailed, in contrast with other studies in which $k_{in}$
was found to be more skewed than $k_{out}$~\cite{Wu2011}.  These
differences of our distributions with other published reports may be
due to the fact that our relatively small user list corresponds to a
subset of users with a high number of retweets in The Guardian's riot
tweet database, and may not be representative of the wider network of
all Twitter users.

\section{Interest communities at different levels of resolution in the Twitter network of the 2011 UK riots}

As a relevant application of current interest, we use the Markov
Stability framework to analyse a directed graph derived from a social
network: the Twitter network derived from The Guardian list of
influential users during the UK riots.

\begin{figure}[t]
\begin{center}
  \includegraphics[width=0.8\textwidth]{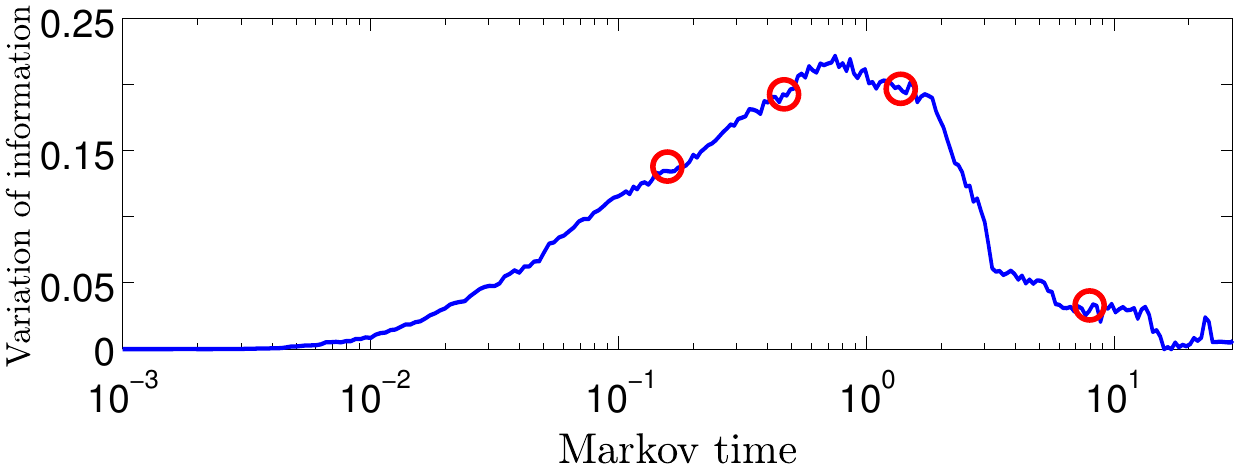}
\end{center}
\caption{
Variation of information of 100 Louvain optimisations of directed
Markov Stability at each Markov time.  The red circles indicate the
Markov times of the communities presented in the main text.}
\label{fig:stability_output}
\end{figure}

In Fig.~1A of the Main Text we show the number of communities 
and in Fig.~\ref{fig:stability_output}, we
show the variation of information (VI) obtained after optimising stability
for Markov times between $10^{-3}$ and $10^{1.5}$. Below we provide
some examples of communities found at different Markov times (marked
with red circles on the variation of information (VI) in 
Fig.~\ref{fig:stability_output}).  These examples
were highlighted due to the relative robustness of the communities, as
well as a means to showcase the different types of communities found
at different levels of resolution.

In the Supplementary Spreadsheet, we provide a spreadsheet with 
 all the communities found at all Markov times.
The convention we have followed to name the communities is
\mbox{T[Markov time]-C[community number]}.

\subsection{Communities at high level of resolution}

\begin{figure}[tp]
\begin{center}
  \includegraphics[width=0.8\textwidth]{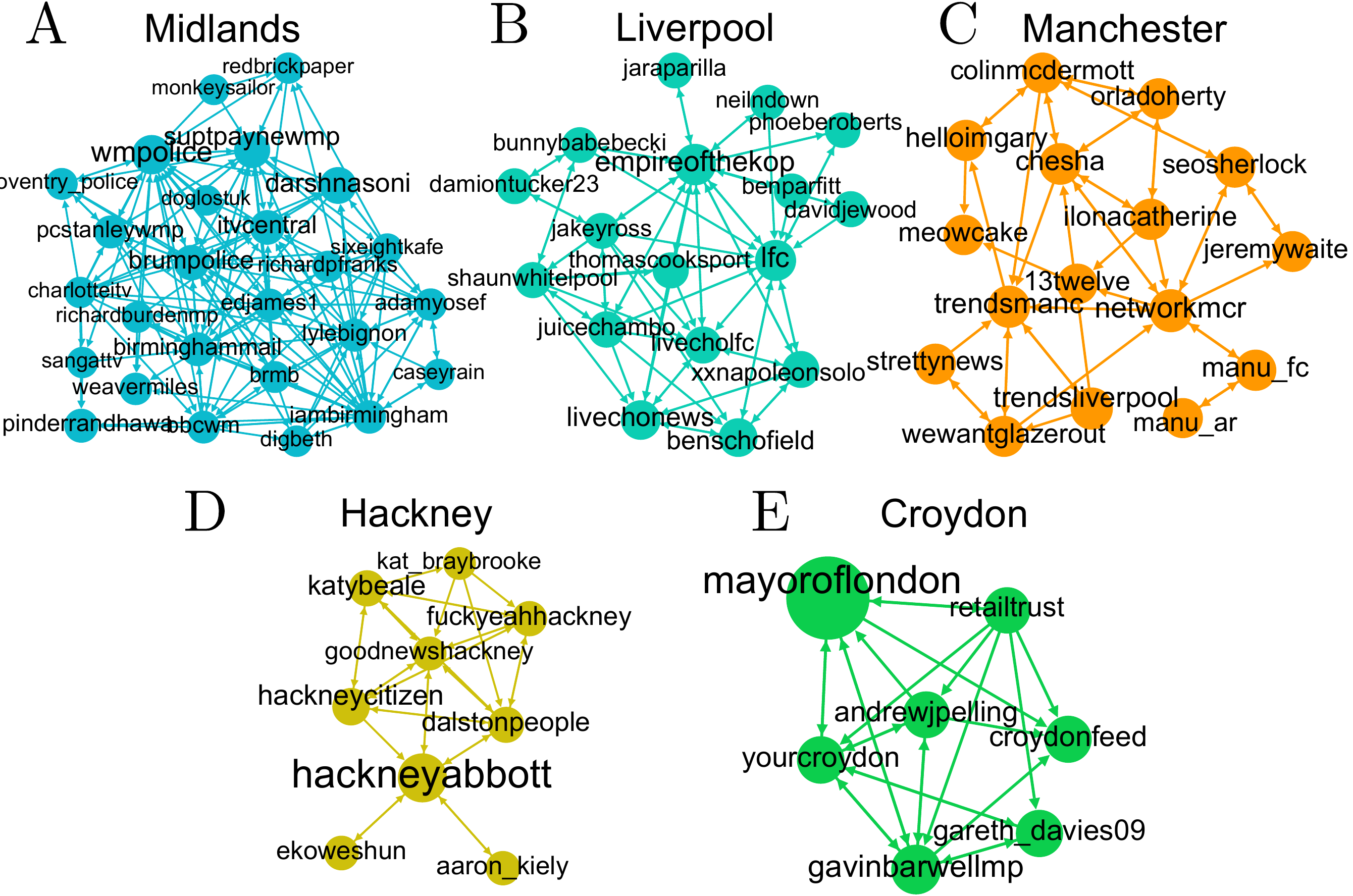}
\end{center}
\caption{Geographical communities found at high level of resolution ($t=0.15$). 
Communities where most members identify
themselves as from the Midlands (A), Liverpool (B), Manchester (C),
Hackney (D), and Croydon~(E).
 }
\label{fig:T0.15-C_local} 
\end{figure}

At short Markov times ($t=0.15$), we find a revealing partition with
149 communities.  Though granular, the community structure of the
partition shows interesting features. Some of the communities in this
partition correspond to a precise geographical location in England or
within a city (Fig.~\ref{fig:T0.15-C_local}).  For example, there are
communities from Hackney (where the riots began) and Croydon. The
latter includes the account of London's Mayor  (mayoroflondon).
Other geographically homogeneous communities are from the Midlands,
Liverpool, and Manchester (T0.15-C10, T0.15-C11 and T0.15-C28).

\begin{figure}[tp]
  \begin{center}
    \includegraphics[width=0.8\textwidth]{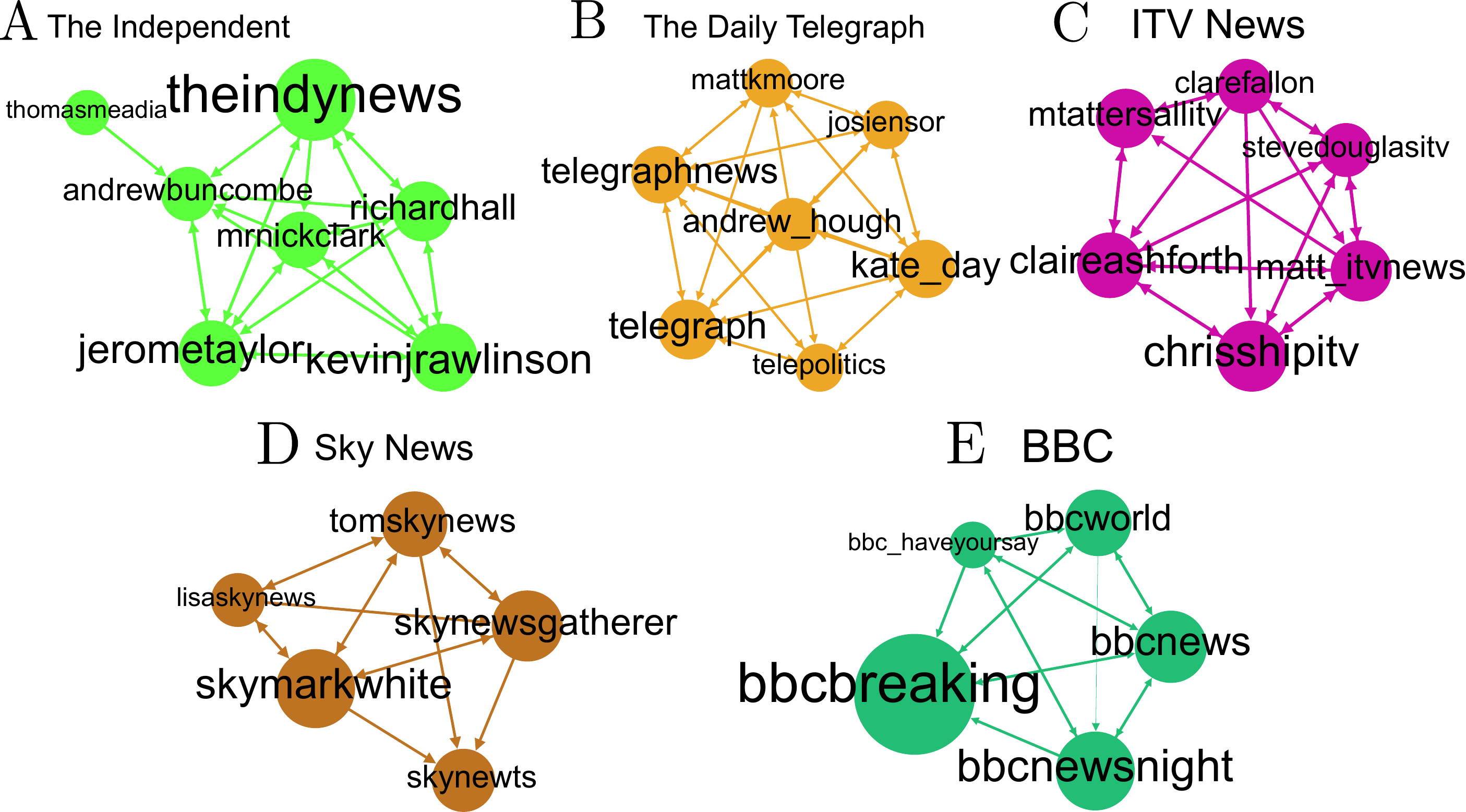}
  \end{center}
\caption{Media communities found at high level of resolution ($t=0.15$): The
Independent (A), The Daily Telegraph (B), ITV (C), Sky News (D), and
BBC (E). }
\label{fig:T0.15-C_media}
\end{figure}

Coexisting with those geographical communities at this level of
resolution, we also find journalists and media outlets in
other communities defined by their affiliation. 
Figure~\ref{fig:T0.15-C_media} shows communities from The
Daily Telegraph, The Independent, ITV, Sky and the BBC. Other
interesting communities in this partition are, for example, formed by
UK activists (T0.15-C0), and another formed by the Anonymous Internet
activist group (T0.15-C12), see the Supplemental Material.

\subsection{Communities at medium level of resolution}

\begin{figure}[tp]
  \begin{center}
    \includegraphics[width=0.8\textwidth]{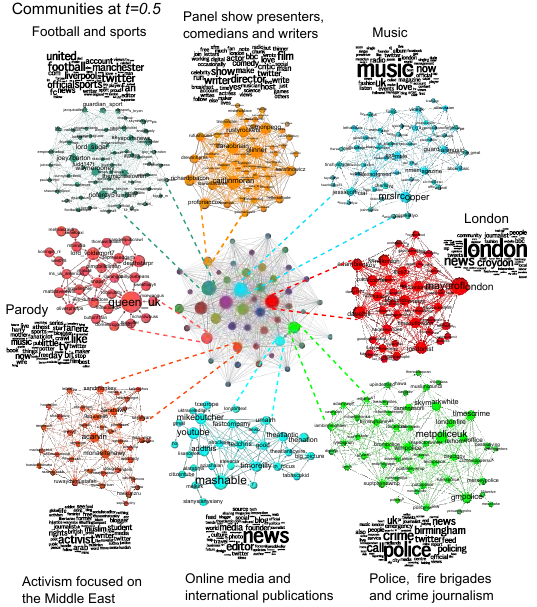}
  \end{center}
\caption{Showcase of communities found at medium resolution ($t=0.5$). At this
Markov time there are 48 communities including one of footballers and
sports-related accounts (T0.51-C0), panel show presenters and
comedians (T0.51-C1), music (T0.51-C4), parody accounts (T0.51-C11),
London (T0.51-C7), activism focused on the Middle East (T0.51-C5),
online media (T0.51-C10), and police forces, fire brigades and crime
journalism (T0.51-C3).
}
\label{fig:T0.5-showcase}
\end{figure}

As the Markov time grows, the communities become coarser. In
Fig.~\ref{fig:T0.5-showcase}, we show the partition of the network at
Markov time $t=0.5$, when there are 48 communities in the riot
network.  The largest community in this partition (T0.5-C0) with 57
members is the `Sports' community.  Other examples of communities
found at this level of resolution include a community of `Comedians,
writers and presenters'; a `Parody' community; a community of `Music
journalists and artists'; a community of `Police forces and crime
journalists'; a `London' community; a community of 'Activist, students
and journalists with a focus on the Middle East'; and an 'Online
media' community of (mostly) Internet media outlets, individuals and
companies.

\subsection{Communities at a coarser level of resolution}

When the Markov time is longer ($t=1.3$) we find a partition into 15
communities in the network.  In Fig.~6 of the Main Text we present a
coarse-grained overview of the communities, their relationships, and
their word cloud self-descriptions.  This partition, its communities,
and the global view of the network it provides are discussed at length
in the Main Text.

\subsection{Communities at low level of resolution}

Finally, at long Markov times ($t=7.4$), we find the partition into 4
communities shown in Fig.~1C of the Main Text.  At this level of
resolution there are three large communities consisting of nodes with
prevailing interest in `Media an entertainment' (T7.4-C0), `UK
politics, activism and journalism' (T7.4-C1), and `International media
and activism' (T7.4-C2).  In stark contrast, 
the fourth community (T7.4-C3) is small and contains only
26 nodes mostly related to the BBC.

\subsection{The change in the community structure when directionality is ignored}
As discussed in the Main text, the community structure detected is
significantly different if the directionality of the edges is
ignored. This phenomenon was exemplified through two examples: the BBC
community (in Fig.~2 of the Main Text, which was heavily affected when
directionality was neglected) and the Monbiot community (in Fig.~3 of
the Main Text, which remained relatively unaffected).

\begin{figure}[t]
  \begin{center}
    \includegraphics[width=0.80\textwidth]{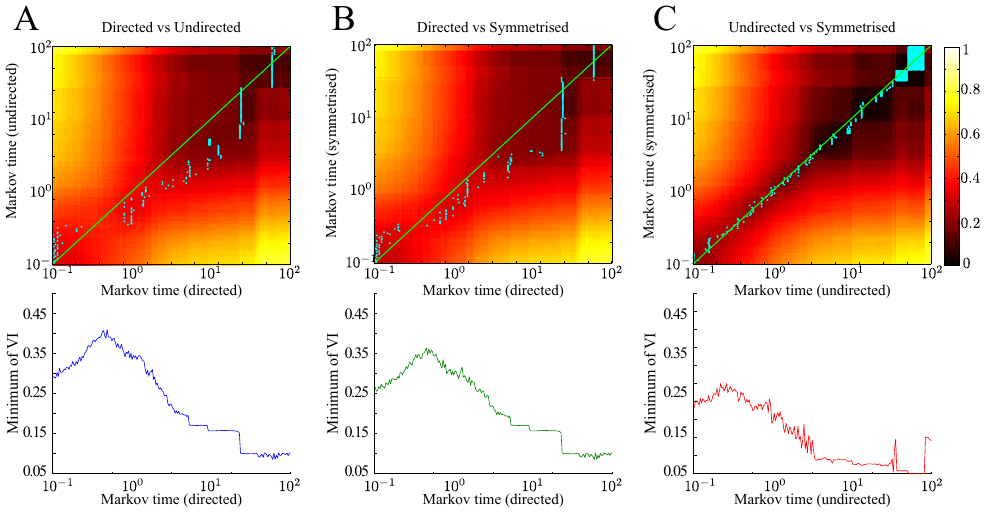}
  \end{center}
\caption{The directed and undirected versions of the Twitter network
  show distinct community structures across Markov times.  {\bf A}:
  Comparison between the directed version of the network and the
  undirected one. Top: Variation of information (VI) between the
  partitions at all Markov times. Low values of VI indicate high
  similarity.  Each row contains the VI between the undirected
  partition to all directed partitions. The minimum of each row is
  shown in light blue. Bottom: These minimum values of the VI from
  each row in the top plot are shown as a function of Markov time. The
  relative high values of the VI indicate that the directed partitions
  are different from the undirected ones across Markov times, but less
  so at large Markov times.  {\bf B}: The comparison between the
  directed network and another undirected version of the same network
  (the $A+A^T$ symmetrised network) shows similar features. {\bf C}:
  The comparison between these two \textit{undirected} versions of the
  network (i.e, the standard undirected and symmetrised networks) show
  that the partitions are highly similar at all Markov times, as can
  be seen by the low values of the VI between the closest partitions,
  and their alignment along the diagonal of the VI plot (top). }
\label{fig:VI_dir_undir}
\end{figure}

To complement this view, we show here the comparison of the
communities found at {\it all} Markov times for the directed riots
network and two undirected versions of it: an undirected network
obtained by simply ignoring the direction of the edges, and a {\it
  symmetrised} version of the network whose adjacency matrix is $A +
A^T$ (i.e., reciprocal edges have twice the weight of nonreciprocal
ones).  In Fig.~\ref{fig:VI_dir_undir}A-B, we show that the partitions
found in the directed network are different from those found in the
undirected and symmetrised versions across all Markov times, whereas
both symmetrised versions are similar to each other
(Fig.~\ref{fig:VI_dir_undir}C).  The differences between the directed
and undirected versions are high at small Markov times and, as
expected, they become smaller as the Markov time grows (i.e., at lower
resolutions). Hence the most prominent effect of ignoring
directionality in this network is to blur the fine structure of the
flow communities. 

For ease of comparison, the full sets of partitions for both the directed and undirected graphs at all Markov times is given in the Supplementary Spreadsheet.

\subsection{Comparison with other community detection methods---Infomap}

As a comparison with our directed Markov Stability methodology, 
we analysed the community structure of the directed Twitter 
network using Infomap~\cite{Rosvall2008, Rosvall2011},
as downloaded from {\tt http://www.mapequation.org/}.
Infomap is a well-known method for community detection in directed
and undirected networks based on information compression, which
has been shown to perform well in some benchmarks with
clique-like communities. 

In this case, Infomap obtains partitions only at two levels of resolution. 
The finer of these two partitions consists of 342 communities: 318 communities
contain only one node; 50 communities contain only two nodes; and the
largest community contains 60 nodes. The coarser of these two partitions has 60
communities: 26 communities still have only one node while the largest
community has 342 nodes, i.e., more than a third of the nodes in the
network.  Hence, in this case, the communities obtained by Infomap lead
to an over-partitioned description of our Twitter network.

We provide all the partitions obtained with Infomap in the
Supplemental Spreadsheet.

These results of Infomap are consistent with the analysis presented in
Refs.~\cite{Schaub2012a,Schaub2012}, where it is shown that Infomap is
a one-step method, which is highly efficient for the detection of
clique-like communities but which may lead to over-partitioning when
the communities are non-clique like. In contrast, Markov Stability
makes use of the full transients (i.e. the complete dynamics with
paths of all lengths, as shown in
Fig.~\ref{fig:riot1000-fixed_output_RBS}A and Fig.~1 from the Main
text) to unfold the community structure across scales.  This tendency
of Infomap to over-partitioning in some networks is signalled by a
large compression gap~\cite{Schaub2012a} with respect to the optimal
compression.  In this particular case, the code-length of the Infomap
partition is 8.3977 bits (after 500 trials)
and the compression gap is 0.3647, which is more than three times
larger than is achieved for clique-like
communities~\cite{Schaub2012a}.  Hence, although in other instances
and benchmarks Infomap performs very well, for this network it does
not produce the nuanced description at different levels of resolution
that our method delivers through the full use of flow
transients~\cite{Schaub2012a,Schaub2012}.  If the direction of the
edges is ignored, this over-partitioning effect of Infomap is even
more striking. For the undirected graph, Infomap obtains two
partitions (code-length: 9.22 bits, compression-gap: 0.455), with 800
communities in the finer partition and 6 communities in the coarser
partition, one of which contains 894 nodes.

\section{Self-descriptions from Twitter biographies}

Interpreting the communities found in the analysis by looking through
all their members is mostly impractical.  As an aid to
assess the quality and intrinsic content of the communities found, we
tap into the information contained in the
mini-biographies provided by the users.  Our premise is that we can
learn valuable information about a community from the small texts that
the members write about themselves. To do this in a more systematic
manner, we collect all the biographies of a community in a single file
(removing urls, emails, numbers, function words, and other nonstandard
characters) and count the occurrences of all words. We then compile the most
frequently used words as an aid in the characterisation of each
community and construct word 
cloud
visualisations~\cite{WordCloudWebsite} of the word-frequencies of the
50 most used words in each community. The word frequencies (and
their word-cloud representation) acts as a `self description' of each
community.

\begin{figure}[tp]
  \begin{center}
    \includegraphics[width=0.80\textwidth]{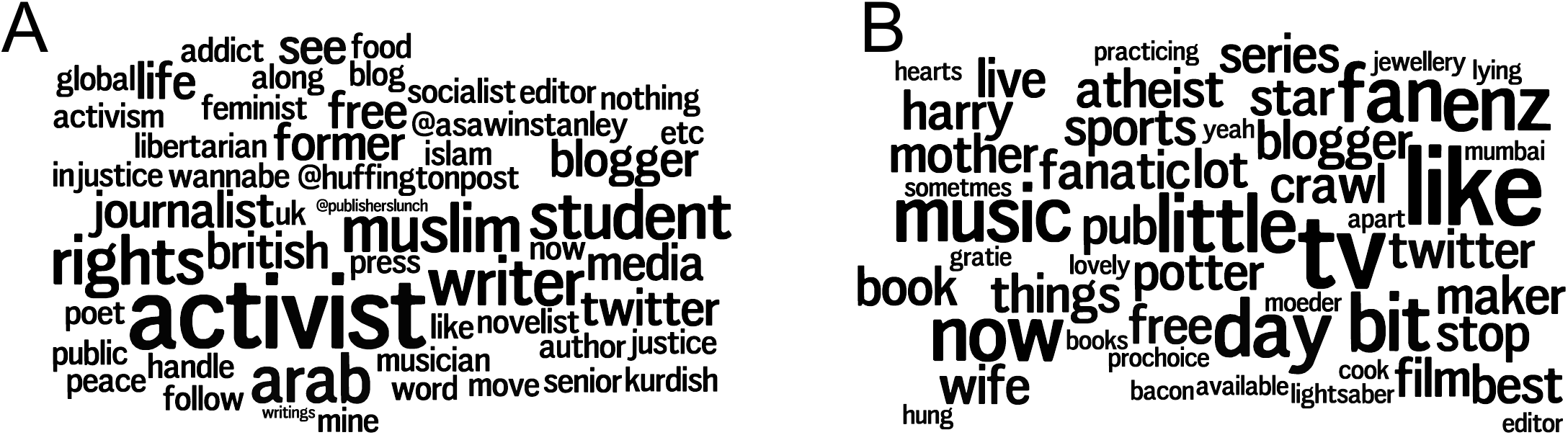}
  \end{center}
\caption{Example word clouds created from the 50 most frequently-used
  words in two communities from Fig.~\ref{fig:T0.5-showcase}.  {\bf
    A}: Word cloud from the biographies of an activist community
  (T0.5-C5). More frequent words appear larger than less-frequent ones
  (function words ignored).  {\bf B}: Word cloud of the Parody account
  community (T0.5-C11) whose members are linked by their interests but
  do not use a common vocabulary to describe themselves
  collectively. In this case the word frequencies do not help
  establish the nature of the community.
}
\label{fig:T0.5-bios}
\end{figure}

In some cases, the self-descriptions of the users in a community share
highly indicative words, which appear prominently in the word clouds.
Figure~\ref{fig:T0.5-showcase} shows that the word clouds of
several communities at $t=0.5$ all represent well their character.
For example, the members of the 'Middle-East activism' community
describe themselves using a consistent vocabulary representing their
interests, as shown in Fig.~\ref{fig:T0.5-bios}A.

On the other hand, other communities are more heterogeneous in the
self-descriptions of their members.  For instance, the members of the
`Parody' community do not use a common vocabulary to describe
themselves (Fig.~\ref{fig:T0.5-bios}B), 
so in this case the word frequencies do not help
establish the nature of the community.  Indeed, this group does not
share a common thematic content but are otherwise linked by their
acting as ironic reflections of a variety of celebrities. This is
reflected in their identification as an interest community.  If we
analyse the membership of the community carefully, one can see that
the community contains many parody accounts (e.g., parodies of the
Queens of England and the Netherlands, and Star Wars and Lord of the
Rings characters).

In general, given the small amount of text available in the
self-descriptions, the word frequencies and word-clouds must be used
judiciously. However, they can be of great aid in providing a simple
visual interpretation of the communities, as shown in the figures in this
Supplemental Information and the Main Text.

\section{Using the RMST-RBS similarity graph to uncover roles in the network}

The RMST-RBS graph thus constructed is a new graph (undirected and
unweighted), which captures geometrically how similar two nodes are,
based on their vectors of incoming and outgoing flow profiles.
Clearly, this role similarity graph is distinct from the original
graph that originated it: two nodes are connected in the role
similarity graph only if they have similar profiles of incoming and
outgoing paths in the Twitter network, \textit{regardless of whether
  they are neighbours in the original network}. Figure~5 in the Main
Text shows the role similarity graph constructed from the Twitter riot
network.

\begin{figure}[tp]
\begin{center}
  \includegraphics[width=0.8\textwidth]{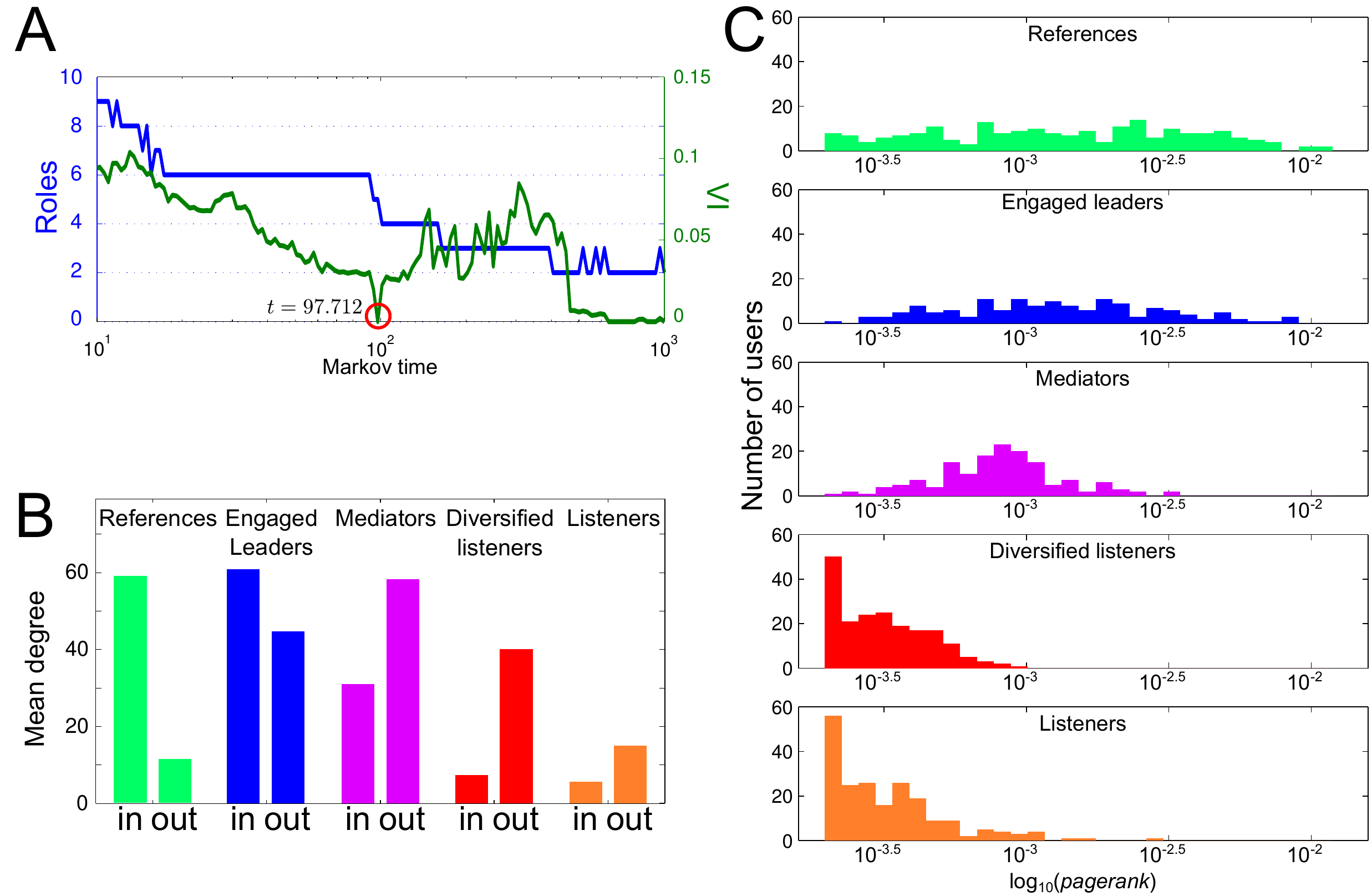}
\end{center}
\caption{{\bf A}: Community detection of the role similarity graph
  using Markov Stability. The blue line shows number of communities
  and the green line shows the variation of information.  A robust
  partition into five communities is detected at Markov time
  $t=97.712$ corresponding to the five roles in the Twitter network.
  {\bf B}: The in- and out-degrees of the five clusters found reflect
  their characterisations as: references, engaged leaders, mediators,
  diversified listeners, and listeners.  
   {\bf C}: Histogram of PageRank values in
  each of the five role classes. The listener categories contain
  mostly nodes with low page rank while the high PageRank values are
  concentrated in the reference and engaged leader categories.}
\label{fig:riot1000-fixed_output_RBS}
\end{figure}

We apply a graph-theoretical community detection method (in this case
undirected Markov Stability) to the role similarity graph to find if
there any significant groups of nodes with similar flow profiles,
without imposing their number or type a priori.
Figure~\ref{fig:riot1000-fixed_output_RBS}A shows that the Markov
Stability analysis of the role similarity graph finds a very robust
partition into five communities of approximately similar sizes at
$t=97.712$. At this Markov time, the variation of information is 0,
which means that in {\it all of} the 100 times we ran the community
detection algorithm we obtained the exact same partition. These five
communities in the role similarity graph correspond to classes (or
types) of roles in the network.  In the Supplemental Material we
provide the full classification of nodes according to these roles.

To interpret the five clusters found by our analysis, we examine
\textit{a posteriori} different characteristics of their members.
Figure~\ref{fig:riot1000-fixed_output_RBS}B shows the mean in and out
degree in the original Twitter network of the nodes in each class
found.  Two of the groups found have higher mean in-degree (i.e.,
Twitter followers) than out-degree, while for the other three groups
the reverse is true (i.e., they follow more than they are followed
by).  If we coarse-grain the original Twitter network lumping together
all the nodes with the same role (as in Fig.~5 of the Main Text), a
striking pattern of connectivity is revealed with classes of nodes
mostly receiving attention (sinks of interest or sources of
information), other classes mostly behaving as sources of interest
(recipients of information) and other classes in between. This leads
to our renaming our role classes as: references, engaged leaders,
mediators, diversified listeners, and listeners (see Main Text for
more details).
Finally, the PageRank distributions shown in
Fig.~\ref{fig:riot1000-fixed_output_RBS}C also help illustrate the
differences between leader, follower, and mediator roles, but would
not be able to discriminate between the roles obtained from our
analysis of the role similarity graph.

\subsection{Comparison to other classic notions of roles in social network analysis}
The notion of node roles in graphs has been studied from different
perspectives, especially in social network analysis. A classic example
is {\it structural equivalence} (SE) in which two nodes have the same
role if they share the same neighbours~\cite{Lorrain1971, Leicht2006},
i.e., if they are swapped the network remains the same. One can
compute how similar (in the SE sense) are two nodes based on the
number of common neighbours.  SE roles are thus based on computing the
number of common immediate neighbours and bear no resemblance to the
flow roles detected via the RMST-RBS approach. In particular, our
approach allows nodes with no common neighbours to have the same role,
counter to the SE definition.

Another classic notion of role in social networks stems from the
theory of {\it regular equivalence} (RE).  RE uses node colorations
(or labellings) to find groups of nodes with the same
role~\cite{White1983, Everett1994, Everett1996}.  Suppose $u$ is a
node in the network with in-neighbours $N_i(u) = \{j:A_{j,u}=1\}$ and
out-neighbours $N_o(u) = \{j:A_{u,j}=1\}$. The colour of $u$ is $C(u)$
and the colour of the in and out-neighbourhoods are $C(N_i(u))$ and
$C(N_o(u))$ respectively. A coloration is said to be (exactly)
regular~\cite{Everett1994} if for any two nodes $u$ and $v$
\begin{equation}
  C(u) = C(v) \, \Rightarrow \left\{
  \begin{array}{ll}
    C(N_i(u)) = C(N_i(v)) & \text{and}\\
    C(N_o(u)) = C(N_o(v)).
  \end{array}
  \right.
  \label{eq:regeq}
\end{equation}
In a regular (or approximately regular) coloration of a network, nodes
with the same colour are said to have the same role~\cite{Everett1996,
  Luczkovich2003}. In its strict sense, the RE definition of role is
combinatorial (and thus lacking robustness in many real-world networks
networks). Furthermore, it is only based on the consideration of the
coloration of immediate neighbourhoods of each node.  Hence it leads
to a very different classification of roles to that obtained with the
RMST-RBS algorithm.

We have obtained the roles of nodes in the riots network obtained
using RE models, and compared them to the RMST-RBS roles obtained
above.  To obtain the RE classes we use two well-known algorithms:
REGE~\cite{Borgatti1993}, which obtains similarities between the nodes
based on RE, and EXCATRE~\cite{Everett1996} which produces a sequence
of regular (or approximately regular) colorations.  We have included
the EXCATRE and REGE partitions in the Supplemental Spreadsheet.

The EXCATRE algorithm applied to the riots network finds only one
non-trivial coloration with 734 roles. (Two trivial colorations are
also obtained: all nodes with the same colour and each node with its
own colour.)  On closer investigation, the roles identified by EXCATRE
correspond trivially to nodes with identical in and out-degree, and
hence it corresponds to considering only immediate neighbourhoods in
the graph (i.e., it would be found by RMST-RBS setting both $\alpha$
and $K_{max}$ equal to 1 so that only paths of length one are
considered).  In our analysis above, we used $\alpha=0.95$ (which
converges at $K_{max}=133$), and the roles obtained by RMST-RBS
incorporate global flow information from the graph.

The REGE algorithm iteratively constructs a similarity matrix between
nodes based on the similarities between their neighbours.  The
similarity matrix is then clustered using hierarchical clustering
techniques to obtain roles~\cite{Luczkovich2003}.  We apply the REGE
algorithm to the riots network (which converges after 6 iterations)
and obtain five clusters (or roles) with 806, 59, 46, 2 and 1 nodes
each. These roles are not informative: the cluster with 59 nodes
contains all the source nodes (in-degree 0); the cluster with 46 nodes
contains all the sink nodes (out-degree 0); while almost all the nodes
in the network fall in the cluster with 806 nodes which lies between
them.  The two small clusters with 2 and 1 nodes lie in slightly
different positions with respect to the sink and node clusters.  As
with EXCATRE, these roles are the result of an analysis based on
immediate neighbourhoods (in and out-paths of length one).

Although RE-based methods and RMST-RBS attempt to find roles in
networks, their conceptual foundation is different.  Both approaches
aim to identify the `types' of nodes that exist in the network
(allowing for the possibility that two nodes on opposite ends of the
network and no common neighbours could be of the same type) but they
use different information to do so. RE-based methods analyse the
similarities between neighbourhoods under colorations, while RMST-RBS
identifies similarities in the transient pattern of long-range flows
though each node. Because of the use of flows at all scales in the
network, from local to global neighbourhoods, the RMST-RBS method
provides a more balanced classification into node classes.  In
addition, RMST-RBS is less sensitive to small changes in connectivity,
and thus more robust for the analysis of realistic networks.  For
example, if we create an edge from a node in the 806-node REGE cluster
to the 59-node cluster of sources (e.g., when a Twitter user decides
to 'follow'' another user), this node would change roles immediately.
Hence in this specific context, a RE-based analysis of the network is
non robust.

\section{Integrating interest communities and roles:  informational organigrams}
Our two analyses (interest communities and user roles) can be
brought together to classify the informational organisation of
communities, as given by the different mix of roles within each
community.  In Fig.~6 of the Main Text we show that the 15
communities obtained at $t=1.3$ can be broadly classified into four
organigrams, which go from a purely broadcast community (of references
and listeners) to other communities that involve dialogue between
engaged leaders, mediators and diversified listeners.

The four organigrams were found by using the mix of roles of each
community (a five-dimensional vector containing
the proportion of nodes in the community that belong to each
role-class) and performing a simple k-means clustering on the communities. 
In other words, each community can be represented by a point inside the unit
cube in $\R^5$, and with k-means we identify the clusters of
communities whose role mixes are similar.

{\tiny

}

\end{document}